\documentclass[journal]{IEEEtran}
\ifCLASSINFOpdf
  \usepackage[pdftex]{graphicx}
\graphicspath{{.}{figures/}}
 \DeclareGraphicsExtensions{.pdf}
\else
\fi

\begin{document}
\title{Development and commissioning of the Timing Counter for the MEG Experiment }

\author{M.~De~Gerone, S.~Dussoni, K.~Fratini, F.~Gatti, R.~Valle, G.~Boca, P.~W.~Cattaneo, R.Nard\`o, 
M.~Rossella, L.~Galli, M.~Grassi, D.~Nicol\`o, Y.~Uchiyama, D.~Zanello
\thanks{M.~De~Gerone, S.~Dussoni, K.~Fratini, F.~Gatti, R.~Valle are with the INFN and University of Genova, 
Via Dodecaneso 33, I-16146 Genova, Italy}%
\thanks{P.~W.~Cattaneo, M.~Rossella are with the INFN of Pavia, 
Via A.~Bassi 6, I-27100, Italy}%
\thanks{G.~Boca, R.Nard\`o are with the INFN and University of Pavia, 
Via A.~Bassi 6, I-27100, Italy}%
\thanks{L.~Galli, M.~Grassi, are with the
INFN of Pisa, Largo B.~Pontecorvo 3, I-56127 Pisa, Italy}%
\thanks{D.~Nicol\`o is with the
INFN and University of Pisa, Largo B. Pontecorvo 3, I-56127 Pisa, Italy}%
\thanks{Y.~Uchiyama is with ICEPP, The University of Tokyo, 7-3-1 Hongo, Bunkyo-ku, Tokyo 113-0033, Japan}%
\thanks{D.~Zanello is with the
INFN of Roma, P.le A.~Moro 2, I-00185 Roma, Italy}%
}

\maketitle
\begin{abstract}
The Timing Counter of the MEG (Mu to Electron Gamma) experiment is designed to 
deliver trigger information and to accurately measure the timing of the $e^+$
in searching for the decay $\mu^+ \rightarrow e^+\gamma$.
It is part of a magnetic spectrometer with the $\mu^+$ decay target in the center.
It consists of two sectors upstream and downstream the target, each one with two layers: the inner one made with scintillating fibers read out
by APDs for trigger and track reconstruction, the outer one consisting in  
scintillating bars read out by PMTs for trigger and time measurement. The design criteria, the obtained performances and the 
commissioning of the detector are presented herein.

\end{abstract}

\begin{IEEEkeywords}
Timing Counter, scintillator counter, high resolution timing detector.
\end{IEEEkeywords}

%
\IEEEpeerreviewmaketitle

\section{Introduction}
\IEEEPARstart{I}{n} the Minimal Standard Model (SM) the Lepton Flavor
Violating (LFV) decays are proportional to $(\frac{m_{\nu}}{M_{W}})^2$, 
(where $m_{\nu}$ is the neutrino mass and $m_{W}$ the $W$ boson mass)
so that the predicted LFV Branching Ratios (BR) are negligible 
($<10^{-40}$).

On the other hand most theories proposed as SM extensions
(e.g. SUSY) predict much larger LFV BRs. 
In \cite{barbieri} the BR($\mu^+ \rightarrow e^+\gamma$) is 
predicted in the range $10^{-12} - 10^{-14}$ 
that is accessible by a high precision experiment, like MEG (Mu to 
Electron and Gamma) \cite{meg2009,meg2011}.

Observation of this decay would be a definitive proof of new 
physics beyond the Standard Model. 
The aim of MEG is to measure this BR
down to a few times $10^{-13}$ improving the current limit of
$1.2\times 10^{-11}$ estimated by the MEGA collaboration \cite{brooks_1999_prd}.

%
%
%
%

\section{Event signature, background and experimental apparatus}
The basic requirement for the MEG experiment is an intense source 
of $\mu^+$, like the one currently available at the Paul Scherrer 
Institute in Villigen (CH), capable of delivering up to $10^8\mu^+/\mathrm{s}$ 
in DC mode \cite{beamline}.
The $\mu^+$s are stopped in a thin target inside a
magnetic spectrometer and decay at rest, hence the $\mu^+ \rightarrow
e^+\gamma$ signature is given by the two-body kinematic, with one photon
and one positron simultaneously emitted back-to-back with the same momentum,
$52.83\,\mathrm{MeV/c} \approx m_{\mu^+} /2$. 
The recognition of the $\mu^+\rightarrow e^+ \gamma$ relies on the precise measurement 
of the kinematic variables that identify the
event: the $e^{+}$ and the $\gamma$ energy and the relative emission angle
and timing.

The major challenge is given by the high
$\mu^+$ stopping rate required to collect enough statistic in a reasonable
time interval, resulting in a large background that has to be carefully
identified and rejected.
There are two major sources of background, in which a positron and a
photon can fake a true $\mu^+\rightarrow e^+
\gamma$ event: the physical or prompt background originating from radiative decay
$\mu^+\rightarrow e^+\nu\bar\nu\gamma$ with the neutrinos carrying small energy, 
and the accidental background arising from coincidences between a high
energy $e^+$ from Michel decay $\mu^+\rightarrow
e^+\nu\bar\nu$ and a high energy photon from other sources like radiative
decay, positron annihilation in flight or bremsstrahlung. 
In the former background source, the time of the $e^+$ and the photon at the
decay vertex are coincident as for the signal, hence the relative timing is
not a variable resolving signal from this background.

In order to minimize the background on the experiment, a detector with
excellent spatial, timing and energy resolutions is required, as  
proposed in \cite{meginfn} (all values quoted are Full Width at Half Maximum, FWHM):
\begin{itemize}
\item Photon energy: 4\%;
\item Positron energy: 0.7$-$0.9\%;
\item Single particle timing (equal for photon and positron): 100~ps;
\item  Photon-Positron relative timing: 150~ps;
\item Photon-Positron relative angle: $17-21~\mathrm{mrad}$.
\end{itemize}
With these performances and the needed $\mu $ stop rate ($10^7- 10^8~\mu/s$) the rate of accidental background  is dominant with respect to the prompt one; in this scenario a high timing resolution helps in rejecting fake events. Overall background rate is evaluated to be $\sim 3\times 10^{-14}$ under these conditions.

\begin{figure}[!ht]
\centering
\includegraphics[width=3.45in]{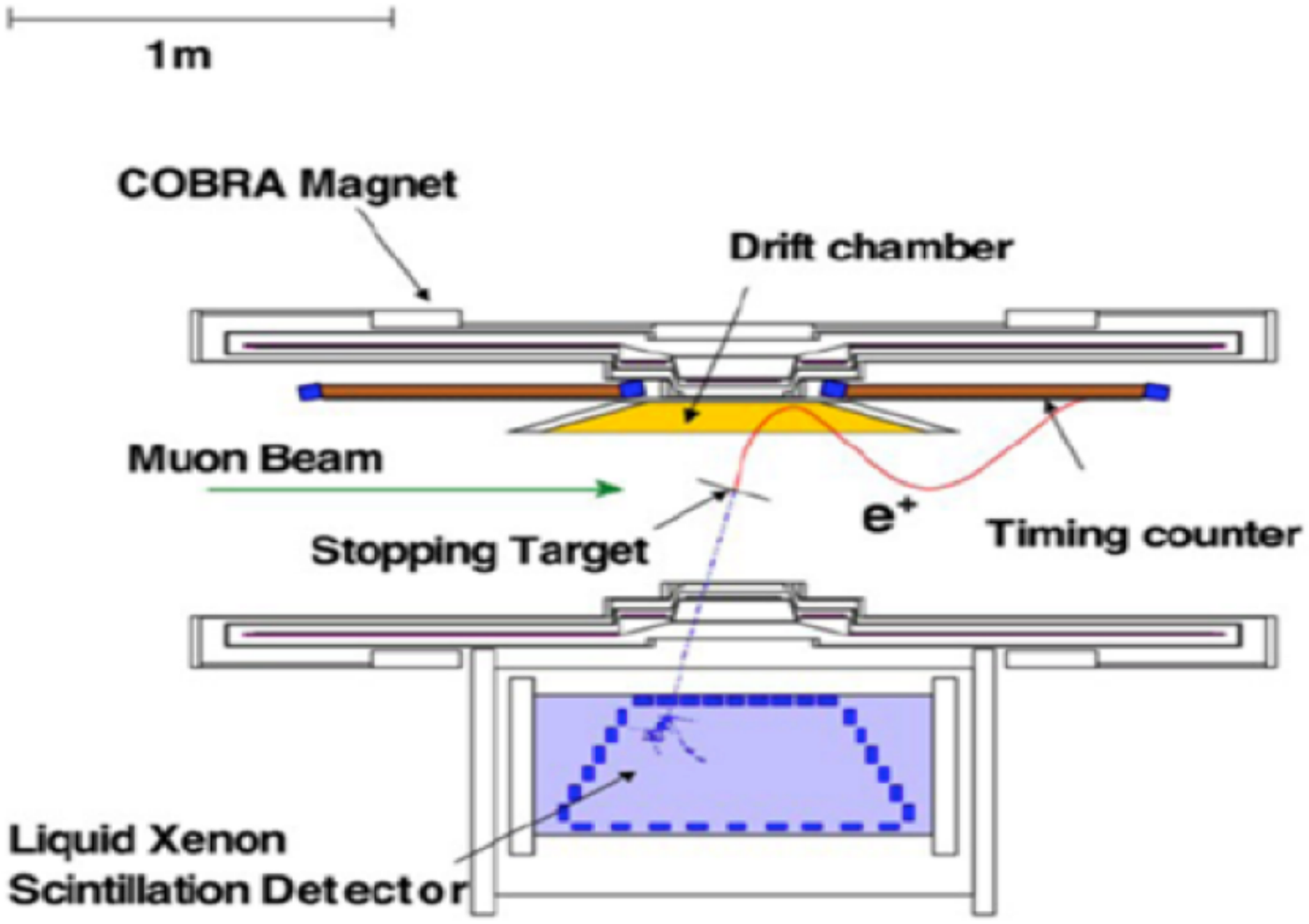}
\caption{Schematic sectional view of the MEG experiment displaying the main components.}
\label{fig_apparato}
\end{figure}

\begin{figure}[!ht]
\centering
\includegraphics[width=3.45in]{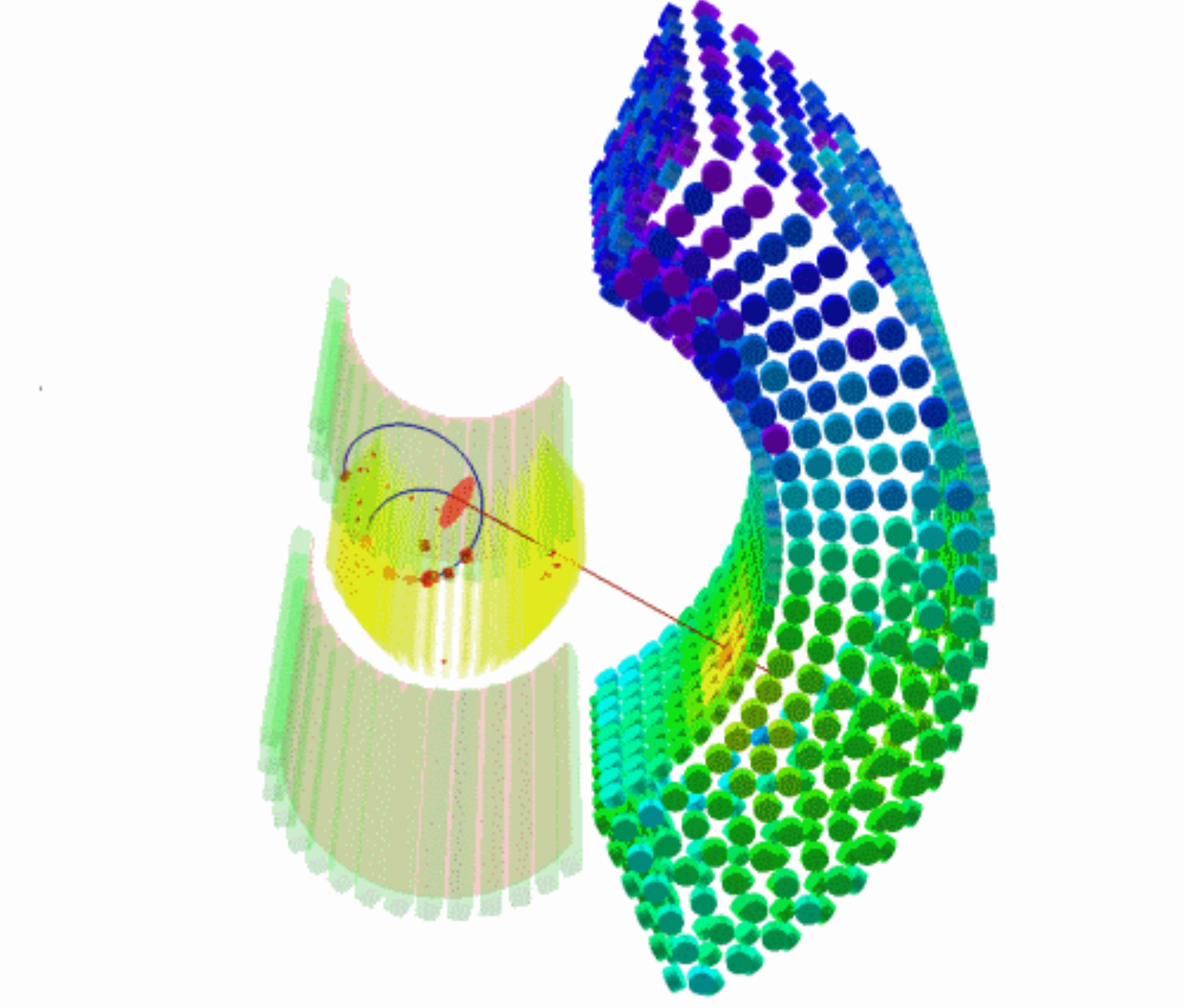}
\caption{3D view of the MEG experiment with a reconstructed candidate event.}
\label{fig_apparato_3d}
\end{figure}

$\mu^+$ decays occur in a $205\,\mu \mathrm{m}$ thick polyethylene target placed
at the center of a quasi solenoidal magnetic field provided by the 
COnstant Bending Radius (COBRA) magnet \cite{ootani_2004}. 
Positrons are at first tracked  by a set of Drift CHambers (DCH),
and then impinge on the Timing Counter (TC)~\cite{tcsiena} that provides position and 
high resolution time information~\cite{megspe}. A schematic drawing of the 
detector is shown in Fig. \ref{fig_apparato} and in Fig.\ref{fig_apparato_3d}.

The particular configuration of the COBRA field enables the $e^+$
bending radius, projected onto the plane transverse to the $z$ axis
\footnote{A cylindrical coordinate system with the $z$ axis along the
  beam is assumed}, 
to be mainly determined by the total momentum with only a mild
dependency on  its polar emission angle ($\Theta$). 
This principle defines an energy threshold for positrons to 
impinge on the TC thus 
strongly reducing  the detector occupancy and the trigger rate from
positrons out of the signal region. 

\begin{figure}[!ht]
\centering
\includegraphics[width=3.45in]{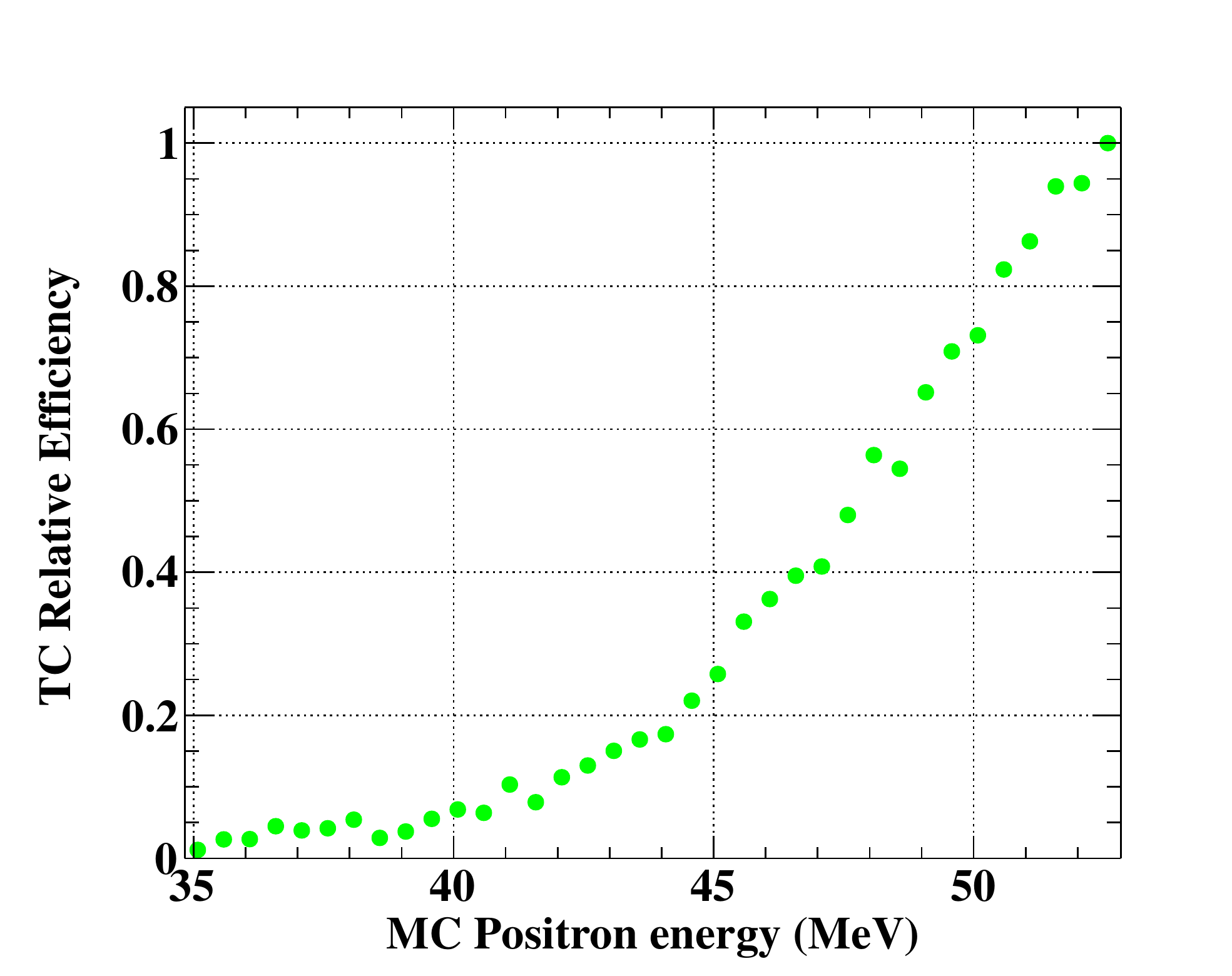}
\caption{Efficiency of the TC versus the $e^+$ energy relative to the signal energy ($52.83\,\mathrm{MeV}$).}
\label{tcreleff}
\end{figure}

Fig.\ref{tcreleff} shows that the threshold effect is somehow blurred 
by the spread in the decay points in the target and the secondary 
interactions in the detector material, leading to lower energy
positrons passing the cut and contributing to detector
crowding. Nevertheless the positron TC 
efficiency drops of a factor 2 (10) for positrons 
$\sim 5.0\,\mathrm{MeV}$ ($\sim 10.0\,\mathrm{MeV}$) below the signal
energy. 

Photon energy, timing and direction are reconstructed in the Liquid
XEnon Calorimeter (LXEC) \cite{megxec}, a vessel, located outside the magnet, 
containing $\sim900$ liter of liquid Xenon whose scintillation 
light is read by 846 Photomultipliers (PMTs). The geometrical acceptance
of the LXEC covers approximately the solid angle delimited in azimuth by 
$120^\circ< \phi <240^\circ$ and $|\cos \Theta|<0.35$ for $\mu^+$ decaying
at the center of the target.

Most of the detector signals are digitized by fast ADCs based on the Domino Ring
Sampler (DRS) chip: the longitudinal TC and the LXEC sampling frequencies are set
at $1.6\,\mathrm{GHz}$ to preserve the time information. 
The slower signals from the DCH are digitized at $800\,\mathrm{MHz}$
\cite{ritt_2010}. The DRS incorporates on a single chip 
an innovative circular sample-and-hold to continuously sample the signals. When a trigger is
fired the sampled waveforms are digitized and stored.

The trigger system digitizes signals from all detectors using a set of
$100\,\mathrm{MHz}$, 10-bit flash 
ADCs and process them with dedicated FPGA (Field Programmable Gate Array): 
this flexible scheme allows to implement many decision algorithms to
quickly select interesting events. The one dedicated to candidate
$\mu^+\rightarrow e^+\gamma$ events is based on fast time and
direction information provided  by the longitudinal TC and LXEC PMTs.

A set of calibrations have been developed for each detector to 
correct for time offsets, space misalignment and energy scale. 
In the following chapter the TC design and calibrations will be described 
in detail, with a summary of the performances during last year run.

\section{The Timing Counter concept and design}
The TC provides a precise measurement of the positron timing and impact
point, to be used for the event reconstruction, and the direction and timing
information of the positron, for a fast $\mu^+\to e^+\gamma$ trigger.
This information must be provided over the full geometrical acceptance
for signal events with the photon pointing to the LXEC inner face. 

The angular range of the positron acceptance cannot be defined easily
in terms of the LXEC geometry because  of the spread of the decay
vertex on the target, in particular along $z$. The angular range to be
covered by the TC is therefore determined by Monte Carlo (MC) simulations
taking into account the mechanical constraints. 

The MC simulations have also been used to determine the TC
radial position to ensure the full efficiency for the signal
simultaneously limiting the fraction of positrons from 
Michel decays hitting the TC that contribute to the accidental trigger
rate. 

\begin{figure}[!ht]
\centering
\includegraphics[width=0.4\textwidth,angle=90]{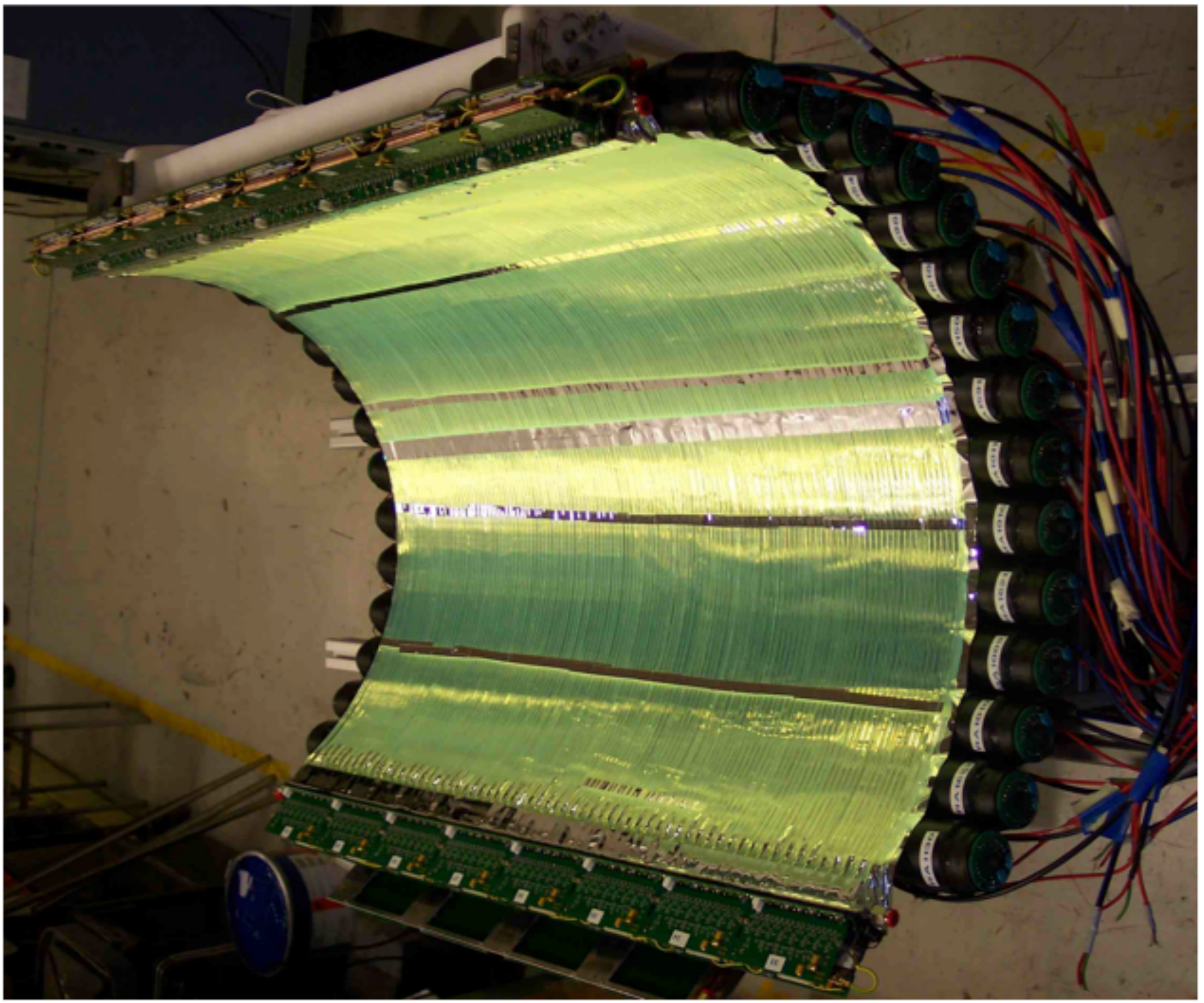}
\caption{Picture of a TC sector: on top the fibers connected to the
  APD read out equipment, below the bars connected to  the PMT. }
\label{TC}
\end{figure}

The detector consists of 2 identical sectors, placed symmetrically
with respect to the target in order to cover the corresponding LXEC
geometrical acceptance. Each sector is divided in two sub-detectors
based on fast optical devices arranged in two layers: the longitudinal
(outer) one segmented along the $\phi$ coordinate and the transverse (inner)
one measuring the $z$ coordinate, satisfying different requirements
(Fig.~\ref{TC}). 

\subsection{The Longitudinal ($\phi$) Detector}
The longitudinal detector is made of an array of 15 scintillating bars
(Bicron BC404) with roughly square shape and dimensions $4.0\times
4.0\times 79.6\,\mathrm{cm}^3$; each bar is read-out by a couple of 
Fine Mesh PMTs (Hamamatsu R5924) glued at the ends. 
The bars are arranged in a barrel-like shape to fit the COBRA magnet 
profile with a $10.5^\circ$ gap between adjacent bars. The choice of
scintillator type, PMT and bar  sizes has been discussed in detail
in~\cite{bonesini_2006,dussoni_2010}.  

\begin{figure}[!ht]
\centering
\includegraphics[width=3.4in]{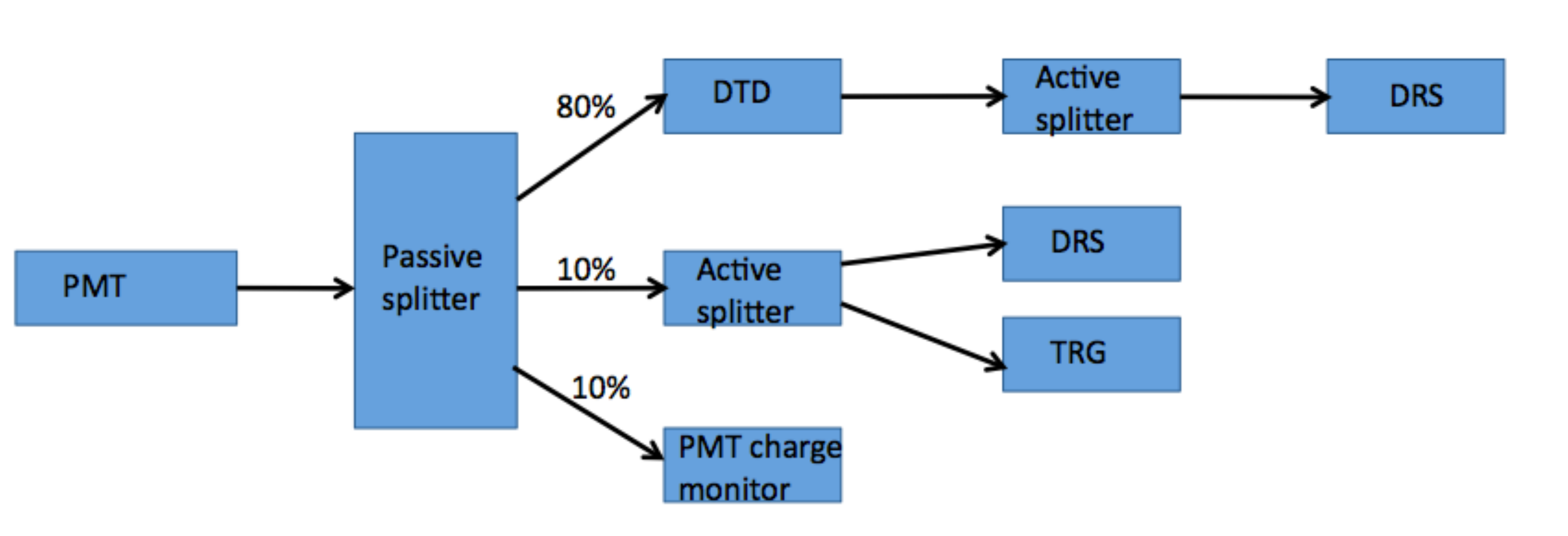}
\caption{Schematic picture of the TC electronic read-out. Signals from each PMT are passively split and fed into dedicated electronics channels. In order to optimize time and pulse reconstruction performances different splitting ratios have been used.}
\label{DTD}
\end{figure}

A schematic picture of the electronic readout is shown in
Fig.~\ref{DTD}. Each PMT signal is passively split in three channels with 80\%, 10\%\
and 10\%\ amplitude fractions, respectively.
The highest fraction is sent to the Double Threshold Discriminator (DTD)
specifically designed for time analysis purposes.

DTD is a high bandwidth, low noise discriminator with two
different tunable thresholds. The first threshold is set as low as
possible, compatibly with the noise level, in order
to minimize the Time Walk effect (see SubSect.\ref{sec_timewalk}). The
second threshold is set to accept only signals associated to physical
events, rejecting spurious hits with low energy deposit.  
The value of the high threshold determines the detector
efficiency and is related also to the trigger threshold as
will be discussed in Sect.\ref{sec_DTD_eff}. When the DTD is fired a
standard NIM waveform is generated and then digitized by DRS
boards together with the PMT waveform. 
The NIM and PMT waveforms are analyzed offline to extract 
the hit time. Several algorithms can be applied, the 
default current one is based on template fit on the NIM waveform. 
This method allows to reduce the
contribution from the DRS jitter to the overall time resolution.

The first 10\%\ copy of each signal  is duplicated and fed to the
trigger system and DRS: in this stage a second active splitter is used
for level translation needed to cope with the input ranges of
Trigger and DRS boards; the second one is sent to a charge integrator
to monitor the PMT aging.  

The described architecture provides accurate and reliable
information from both NIM and PMT waveforms for each event and
their combination offers adequate information valuable to further
improve overall performances.

\subsection{The Transverse (z) Detector}
The transverse detector consists of an array of 256 scintillating
Multi-Clad fibers (Saint Gobain BCF20) coupled to Avalanche
PhotoDiodes (APD Hamamatsu S8664-55), placed orthogonally with respect
to the scintillating  bars, positioned on top of them and covering the
same surface. 

The small sectional area of the fibers ($5\times 5\,\mathrm{mm}^2$
plus $2\times 0.5\,\mathrm{mm}$ wrapping thickness totaling $6.0\,\mathrm{mm}$
in size along $z$)  fits the required space resolution and perfectly
matches the $5\times 5\,\mathrm{mm}^2$ APD 
sensitive area. In order to comply with the mechanical constraints, 
the fiber ends at the APD side are bent in two different types: one set  has almost
straight termination while the other is ``S'' shaped with
small curvature radii ($\sim 2\, \mathrm{cm}$): this causes a light loss that is recovered by
wrapping the fibers with a high reflectance film \cite{degerone}. This
results in a 1.0 mm spacing between fibers to accommodate for the
wrapping, leading the pitch to 6~mm along $z$.

The APD advantages can be resumed as: small size, insensitivity to
magnetic field and fast response. The main disadvantage is the low
gain, which can reach at most $\sim 10^{3}$ if biased near the
breakdown voltage, and high capacitance, which makes them sensitive to
noise. 

Each APD is closely coupled to a readout electronics performing signal 
amplification and shaping.
A single APD current
pulse is read as voltage across a load resistor and amplified. Bundles
of 8 APDs are mounted on distinct boards carrying amplifying stages, power
supplies, HV bias and ancillary control signal. Due to mechanical constraints, 
the fiber bundle mounted on a single front-end board is interleaved with the
fiber bundle connected to the adjacent board: the resulting assembly, 16
consecutive fibers read-out by two adjacent boards, covers $9.6\times40.0\;\mathrm{cm^2}$.
\begin{figure}[!ht]
\centering
\includegraphics[width=3.4in]{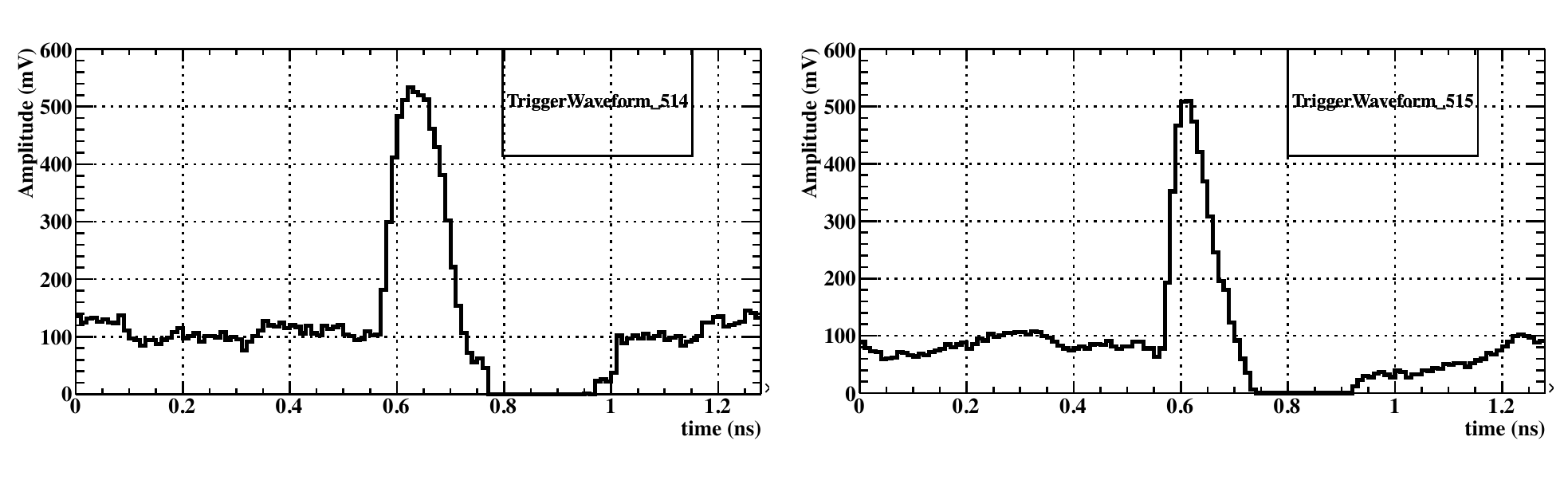}
\caption{Example of a block of 2 trigger waveforms from APDs,
related to interleaved bunch of fibers, and
covering a $\Delta z \approx 9.6~\mathrm{cm}$ wide region on one
(DownStream) sector.   
Each waveform results from the addition of signals from 8 APDs;
vertical scale is 100~$\mathrm{mV}$/div and horizontal scale is
0.2~$\mu\mathrm{s}$/div. A 100~$\mathrm{mV}$  offset is added to keep
the whole signal and baseline into the ADC dynamic range 0-1~$\mathrm{V}$.   } 
\label{APD}
\end{figure}

The analog signals from the 8 APDs on each board are summed and fed
to the trigger digitizer: an example is depicted in Fig.\ref{APD};
this is a block of 16 interleaved fibers read out by 2 consecutive
boards, whose signals can
be summed up in order to increase the Signal-to-Noise ratio and
consequently the trigger reliability. Noise levels are measured to be
in the range 20-30~$\mathrm{mV}$~\emph{r.m.s}. 
Using the analog summed output the impact coordinate $z$ is obtained
at the trigger level with $\mathcal{O}(3\,\mathrm{cm})$ precision. 

\begin{figure}[!ht]
\centering
\includegraphics[width=3.4in]{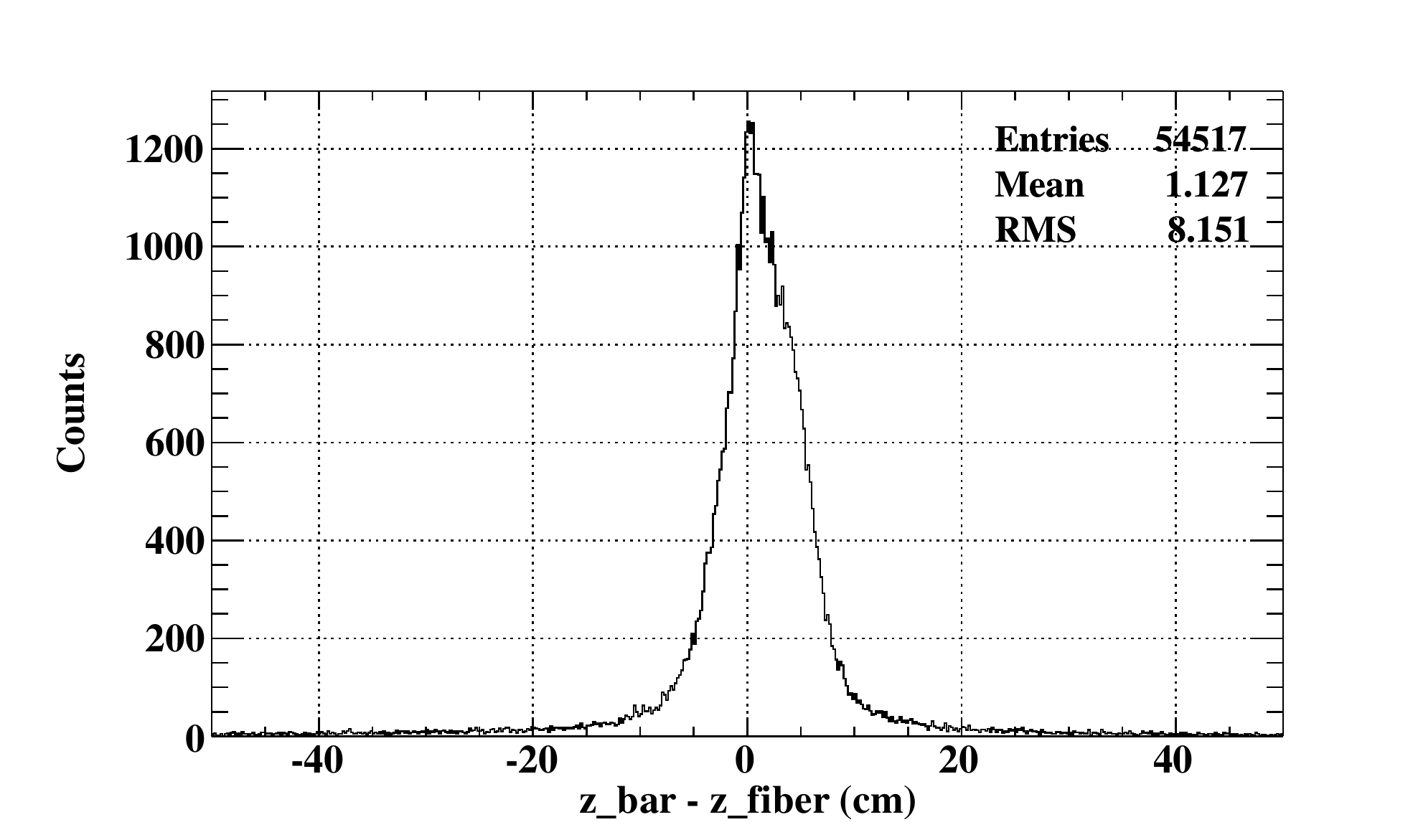}
\caption{Distribution of $(z_{fiber}-z_{bar})$.}
\label{z_APD}
\end{figure}

The front end boards provide also a digital signal from a discriminator
for each APD identifying each single fiber hit. The encoded
information is stored on dedicated FPGA based VME boards. The
coordinate of the fiber $z_{fiber}$ can be matched to the information
obtained from the time difference between opposite PMTs on the same
bar of the longitudinal detector $z_{bar}$.  
An example of a preliminary $z_{fiber}-z_{bar}$ distribution is shown in
Fig.\ref{z_APD}. Since the trigger was fired by cosmic rays hitting the
bars, there are some tails due to the inefficiency of the
APD detector: a $\mathcal{O}$
(10\%) is known to be due to geometrical effects
(incomplete coverage due to mechanical constraints and $1~\mathrm{mm}$
spacing between fibers) but the larger amount comes from inefficient event
reconstruction. A more detailed study on this topic is under way.

\section{The Timing Counter in the trigger}\label{sec_trigger}
%
The trigger system is designed to operate a quick and highly efficient
selection of candidate events, which is strongly needed for an
experiment looking for ultra-rare decays in a huge beam-related
background environment. The severe requirement of a 
maximum latency $\leq 450\,\mathrm{ns}$ 
prevents us from using any piece of information associated with the
DCH, thus leaving the TC as the only detector available for the
positron at the trigger level. The use of the TC in the trigger has a
twofold objective: PMT and APD signals are digitized by means of 8 VME
6U trigger boards (named ``Type1'', 4 for each), each hosting 16 input
channels equipped with 10-bit flash-ADCs at 100~MHz frequency (for
details  on the trigger hardware and architecture refer to
\cite{Luca:2010}). The output data are recorded, and processed, by a
Virtex FPGA to operate an on-line reconstruction of the time and 
impact point of the impinging positrons. Dedicated real-time
algorithms have been developed to implement baseline subtraction and
amplitude calibration to compensate for any PMT gain and bar  
light yield difference.

A TC cluster is defined in association with a positron loosing energy
in a bar such that the signal amplitudes from each PMTs exceed a
pre-defined threshold. In addition, the sum of the two amplitudes must
be higher than another threshold,  which ensures an event selection 
almost independent of the hit position. That threshold is equivalent
to an energy deposit of $\approx 2$~MeV, which makes the selection
fully efficient for signal positrons. If that condition is met by more
consecutive bars (which is the  usual case), these are grouped in a
single cluster where only the pieces of information (time and
position) of the first to be crossed are used. 

The on-line estimator for the positron time $T_{e^+}$ is obtained from
a parabolic interpolation of the  leading edge of the sum of the PMT
signals on the hit bar. A $\approx 2.5$~ns resolution is obtained on
that observable, which allows us to define a conservative 20-ns
coincidence window with the $\gamma$ in the LXEC detector. 

The $z$-coordinate reconstruction proceeds in two parallel streams to
provide, in combination with the bar index, a redundant stereo
read-out of the positron hit position on the TC: 
\begin{itemize}
\item APD signal discrimination (an example of which is shown in
  Fig.~\ref{APD} as recorded by a trigger board); 
\item the use of a look-up table to compute the logarithm of the PMT
  charge ratio which, according to  
Eq.~\ref{qratio}, is linearly correlated to the z-coordinate. 
\end{itemize}
In the former case, due to the fan-in of APD signals (8 APDs to 1
trigger input channel), a resolution $\sigma_z  
\approx 3$~cm is expected. In the latter, the resolution has been
measured to be slightly worse, $\sigma_z  
\approx 5$~cm, but still suited to obtain an efficient background
rejection. 

\begin{figure}[htb]
\begin{center}
\includegraphics[width=3in]{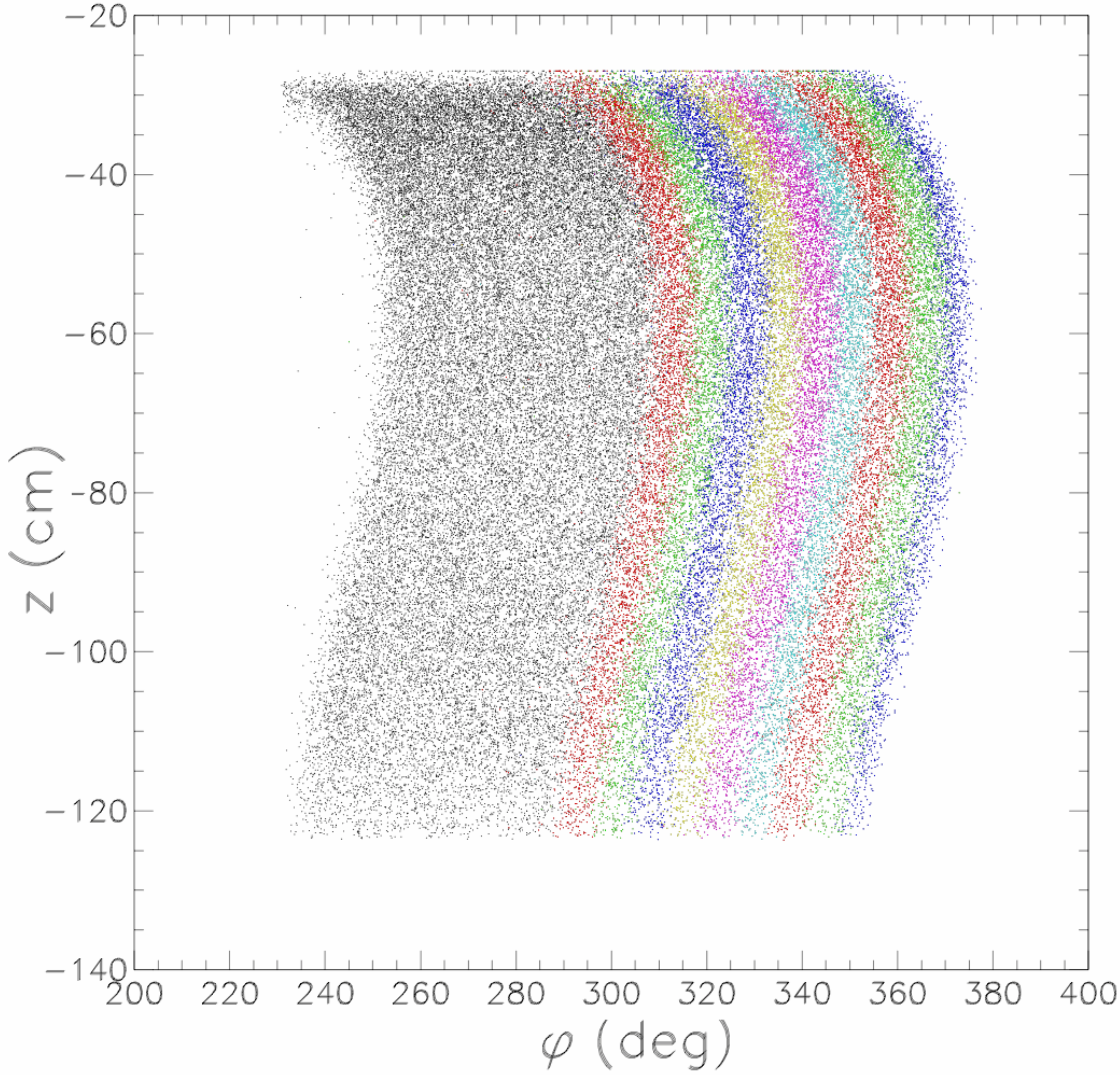}
\caption{Expected hit map $(z,\varphi)$ for positron signals on the TC obtained from a Monte Carlo simulation. Different
color markers refer to positrons emitted at different $\phi$ angles in the detector frame, each band being associated 
with a PMT row in the LXEC.}
\label{fig:match}
\end{center}
\end{figure}

The selection of $e^+ - \gamma$ pairs proceeds as follows. in the
absence of DCH pieces of information, we need to assume that each
positron is a signal-like one: meaning, it must be back-to-back
with the $\gamma$ (whose direction is extracted from the 
position of the PMT in LXEC with the maximum output) and carrying the
signal momentum. Under this assumption, the association of a hit
position on the TC with a photon entering the LXEC with a given
direction is univocal and 
it is straightforward to build an association between the index of the
maximum amplitude PMT in LXEC and the TC bar and APD indices based on
Monte Carlo simulated signal events, as shown in Fig.\ref{fig:match}. This
mapping is summarized in a look-up table which is implemented on-line
to check the back-to-back condition. 

\section{Timing measurement}\label{sec_timgen}
Considering a positron impinging on a TC bar, the time measured by the inner (${in}$) and outer (${out}$) PMTs can be written as:

\begin{eqnarray}\label{tcinouttime}
t_{in}=T_{TC}+b_{in}+TW_{in}+\frac{\frac{L}{2}+z}{v_{\mathrm{eff}}} \nonumber
\\
t_{out}=T_{TC}+b_{out}+TW_{out}+\frac{\frac{L}{2}-z}{v_{\mathrm{eff}}}
\end{eqnarray}
where $T_{TC}$ is the actual time of impact of the positron on the
bar, $b_{in,out}$ are offsets depending on the readout chain and fixed
for each couple of PMTs, $TW_{in,out}$ are contributions from Time
Walk effect, $v_{\mathrm{eff}}$ is the effective velocity of light in the bar
and $L$ is the bar length; the $z$ axis points along the main axis 
of the bar and its origin is taken in the middle of the bar.

From Eq.~\ref{tcinouttime} the true impact time is given by:

\begin{equation} \label{tctime}
T_{TC}=\frac {t_{in}+t_{out}}{2} -  \frac{b_{in}+b_{out}}{2} - \frac{TW_{in}+TW_{out}}{2} - \frac{L}{2v_{\mathrm{eff}}}.
\end{equation}

The determination of the factors $b$ and $TW$ and their stability through the whole data taking period will be discussed hereinafter in more details.

\section{Position measurement}\label{sec_posgen}
The positron impact point is obtained from
the time difference between the two pulses as
\begin{equation} \label{z_time}
z=\frac{v_{\mathrm{eff}}}{2}\cdot (t_{in} - t_{out} - (b_{in} - b_{out}) - 
  (TW_{in} - TW_{out}))
\end{equation}

The impact point can also be evaluated using the ratio between the 
charges delivered at each PMT \cite{knoll}: 
\begin{eqnarray} \label{qq}
Q_{in} = E \cdot G_{in} \cdot
\exp^{-\frac{\frac{L}{2}+z}{\Lambda_{\mathrm{eff}}}} 
\\
Q_{out} = E \cdot G_{out} \cdot
\exp^{-\frac{\frac{L}{2}-z}{\Lambda_{\mathrm{eff}}}} 
\label{eq:qfrac}
\end{eqnarray}
where $E$ is the energy released inside the bar, $G_{in,out}$ takes
into account several contributions (i.e. the scintillator yield, 
PMT quantum efficiency and gain), $\Lambda_{\mathrm{eff}}$ is the
effective attenuation length of the bar. Taking the ratio we obtain: 
\begin{equation} \label{qratio}
\frac{Q_{in}}{Q_{out}}=\frac{G_{in}}{G_{out}} \cdot
\exp^{-\frac{2z}{\Lambda_{\mathrm{eff}}}} 
\end{equation}
which leads to:
\begin{equation} \label{z_charge}
z=\frac{\Lambda_{\mathrm{eff}}}{2}\left(\ln\frac{Q_{out}}{Q_{in}} -
  \ln\frac{G_{out}}{G_{in}}  \right) 
\end{equation}

Moreover, from Eq.~\ref{qq} the energy release in the bar can be
estimated independently from $z$: 

\begin{figure}[!ht]
\centering
\includegraphics[width=3.4in]{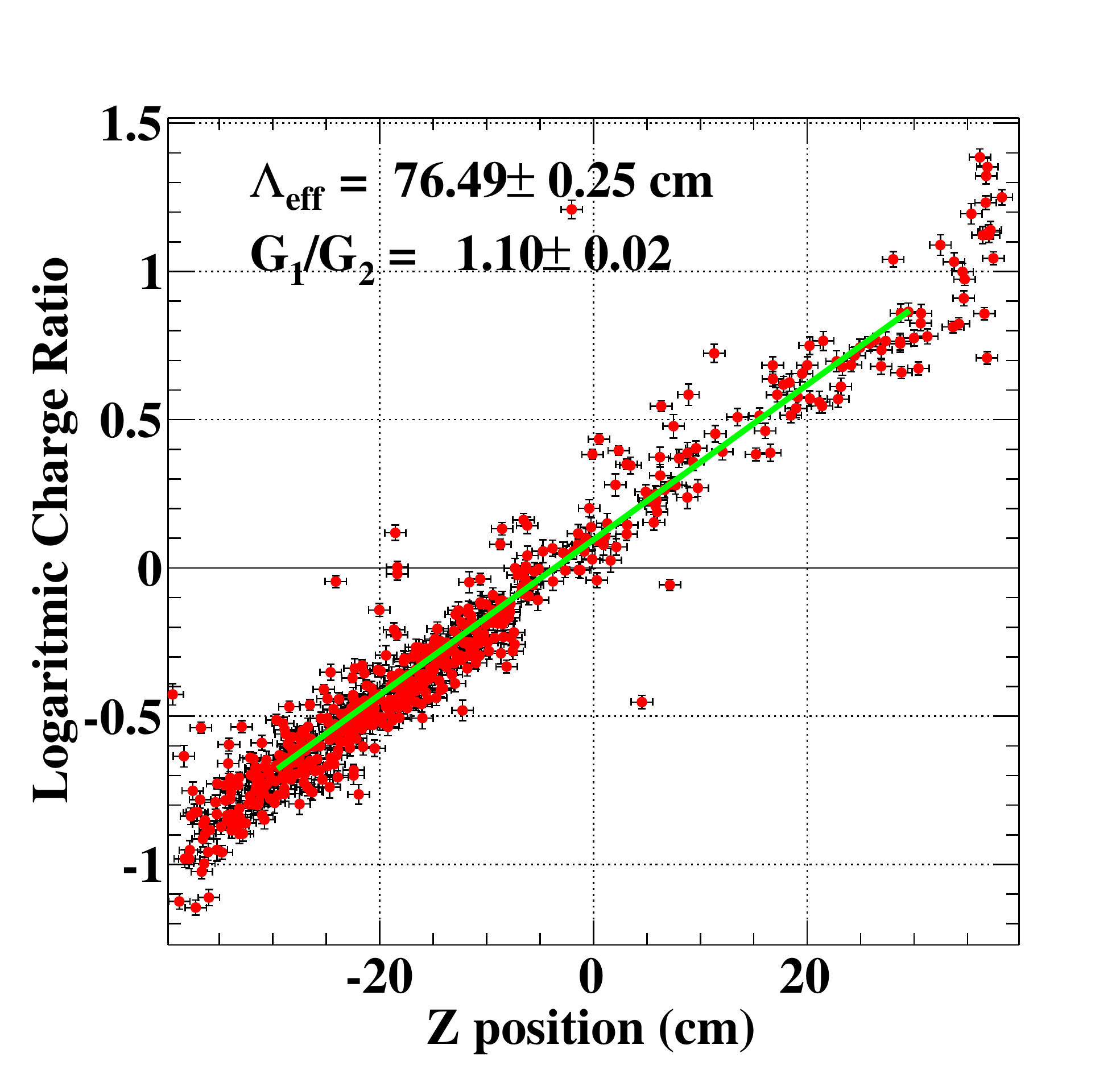}
\caption{Logarithm of the PMT charge ratio vs the $z$ coordinate
  measurement using Eq.~\ref{z_time} for bar number 1. 
The linear fit returns $\Lambda_{\mathrm{eff}}$ and the ratio of PMT
gains.} 
\label{lamqq}
\end{figure}

\begin{equation} \label{energy}
\sqrt{Q_{in}\cdot Q_{out}} = E\cdot\sqrt{G_{out}\cdot G_{in}}
\exp^{-\frac{L}{\Lambda_{\mathrm{eff}}}} 
\end{equation}

Note that the combination of Eqs.~\ref{z_charge} and \ref{energy}
provides a way to evaluate the ratio between
$\Lambda_{\mathrm{eff}}/v_{\mathrm{eff}}$ for each TC bar.  Assuming
$v_{\mathrm{eff}} = 14.0 \mathrm{cm/ns}$, the value of
$\Lambda_{\mathrm{eff}}$ is extracted from a linear  fit as shown in
Fig.\ref{lamqq}. The $\Lambda_{\mathrm{eff}}$s extracted in 2010 for
all bars fall in the range  $40-90\,\mathrm{cm}$.

Two methods for the impact point reconstruction are used in
different stages of the data acquisition chain. The on-line algorithm
for $z$ reconstruction in trigger, requiring fast response and 
moderate precision, relies on the charge ratio method, while the
offline analysis, aiming at the best possible resolution, exploits
the PMTs time difference. The need for offline calibrations to guarantee
the ultimate performance (both for $z$ and time resolutions)
is satisfied by using several tools, presented in the following.

\section{Beam Test performances}
All TC single modules (bar-PMT assembly) were tested using narrow
particle beams to evaluate the timing resolution under conditions
close to the experimental ones. The tests were carried on at the Beam
Test Facility (BTF) in the INFN Frascati National Laboratory. The BTF
can provide bunches of $e^-$ or $e^+$ with energy up to 
$750\,\mathrm{MeV}$ and multiplicity tunable in the 
range $1-10^{10} $ particles/bunch \cite{btfcomm,btf-ieee-2004}. 

In our test, we used it as a single electron source in order to
simulate the bar crossing by one particle at once. The single
bar rate under experimental conditions has been evaluated to be
$\mathcal{O}(100\,\mathrm{kHz})$ by means of MC simulations. Since the
PMT response is as short as $10\,\mathrm{ns}$, there is a 
negligible probability to have overlapping by multiple hits, thus we 
correctly reproduced the actual working conditions.  Rate effects have
been however studied to make sure the detector performance
is not affected by particle crowding and are reported
elsewhere~\cite{bonesini_2006}.

The signals from PMTs were fed into the DTD and then used to provide
Start-Stop signals to a TAC (Time-to-Amplitude Converter) system that
together with an ADC board allowed to obtain a $12.5\;\mathrm{ps/bin}$
resolution, calibrated using high-precision delay lines.
The timing resolution was evaluated from $\Delta T = \frac{t_{in} -
t_{out}}{2}$ distributions taken in various positions, where $t_{in}$ and 
$t_{out}$ are defined in Eq.~\ref{tcinouttime}. This method is almost equivalent to the one making use of impact time from Eq.~\ref{tctime} 
with the following differences: {\em i)} $\Delta T$ is not affected by
Time Walk effect at the leading order approximation and {\em ii)} a
significant contribution to time spread can be due to the finite spot
size. The Time Walk treatment under experimental conditions is discussed in
\ref{sec_timewalk}, while the beam spot size contribution was evaluated as follows:
the impact point position was determined with an
external telescope defining a $5\,\mathrm{mm}$ spot on the bar under test. The effective
velocity of light in the bar was estimated from $\Delta T $ vs $z$, 
to be $v_{\mathrm{eff}} \sim 14.0\,\mathrm{cm/ns}$.
The resulting contribution from the impact point uncertainty to the measured timing
resolution is given by $\sigma(\Delta T) = 0.5/\sqrt{12}\,\mathrm{cm} / (v_{\mathrm{eff}}
(\mathrm{cm/ps})) = \mathcal{O}(10\,\mathrm{ps})$
under the approximation of a uniform beam spot (worst case), and thus
negligible. A similar contribution was evaluated
for the electronic chain (DTD-TDC-ADC) by splitting the signal from one
PMT and looking to the width of $\mathrm{T_{stop}}-\mathrm{T_{start}}$ distribution.
A mechanical support allowed scanning each bar by moving the impact
point along its main axis ($z$ direction) and varying the impact angle to mimic 
the experiment conditions. A rear detector was used in order
to veto multi-particle bunches that could affect the time measurement.

\begin{figure}[!ht]
\centering
\includegraphics[width=3.4in]{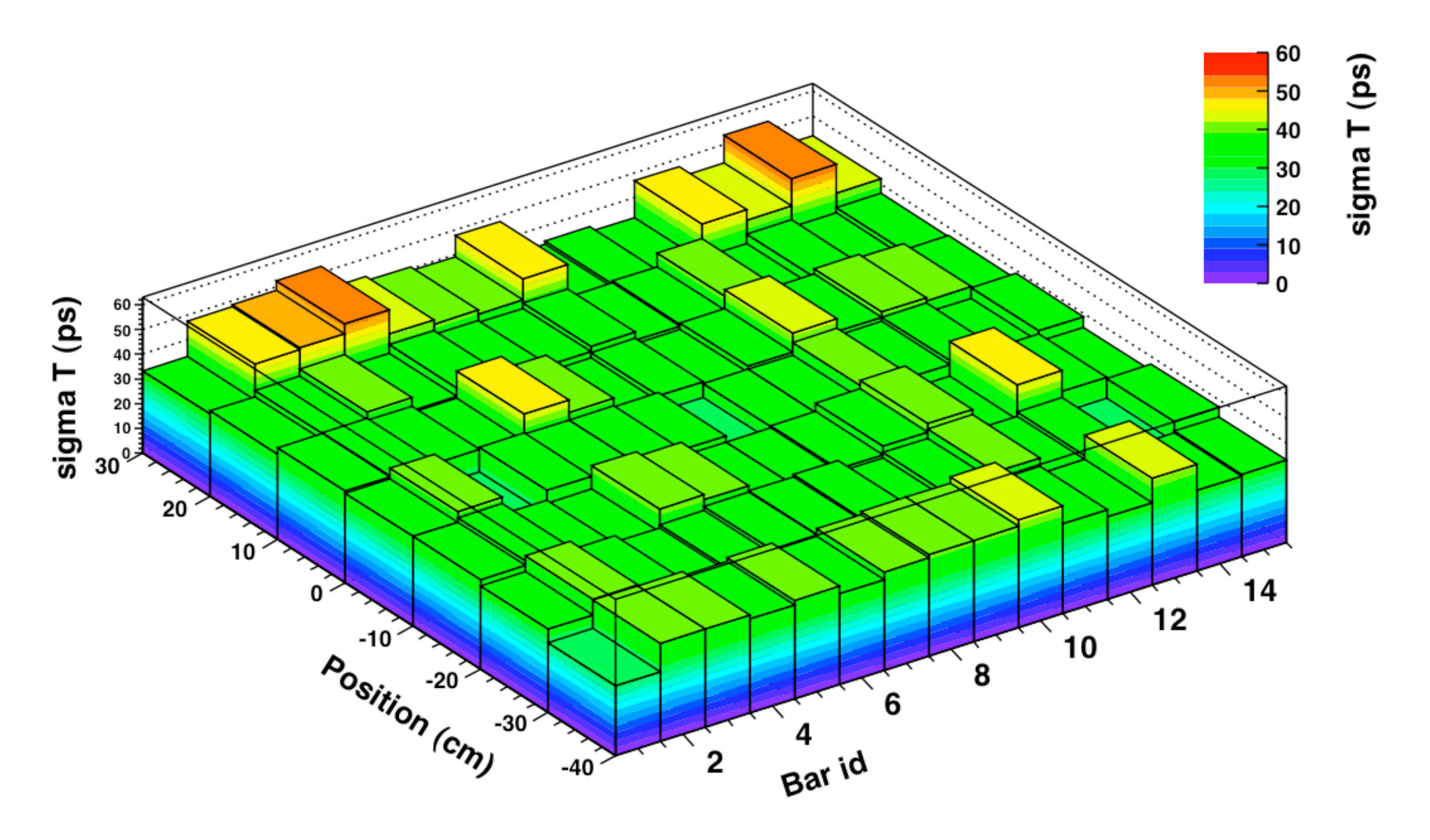}
\caption{Bar time resolution, as a function of bar number and impact point position.}
\label{figbtf}
\end{figure}

\begin{figure}[!ht]
\centering
\includegraphics[width=3.0in]{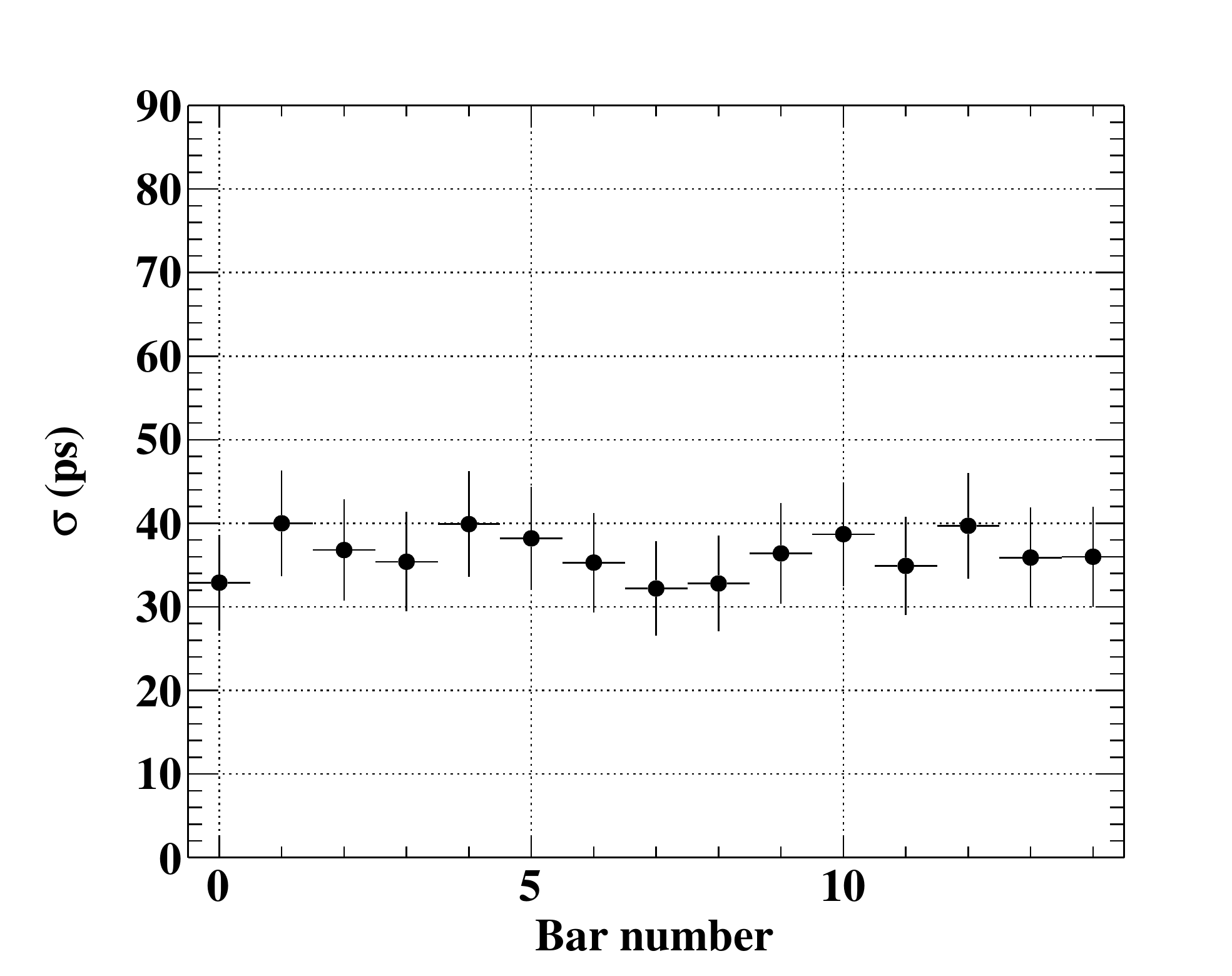}
\caption{Bar time resolution averaged along the bar length, versus bar number.}
\label{figbtf_media}
\end{figure}

A r\'esum\'e of the measured resolutions is presented in fig.~\ref{figbtf} and
in fig.~\ref{figbtf_media} from which it is clear that the averaged time 
resolution for each bar is $\sigma_{T} \leq 40\,\mathrm{ps}$.

\section{Calibrations}
The integration of the TC in the MEG apparatus requires a full set of
calibrations to determine the detector parameters
necessary to optimize its performances. 
Hereinafter only those calibrations involving the TC alone are described. 
Additional calibrations, involving other sub-detectors (e.g. space and time 
alignments with DCH and LXEC) are described elsewhere~\cite{tccalib}.

\subsection{PMT gain equalization}
\begin{figure}[!ht]
\centering
\includegraphics[width=3.4in]{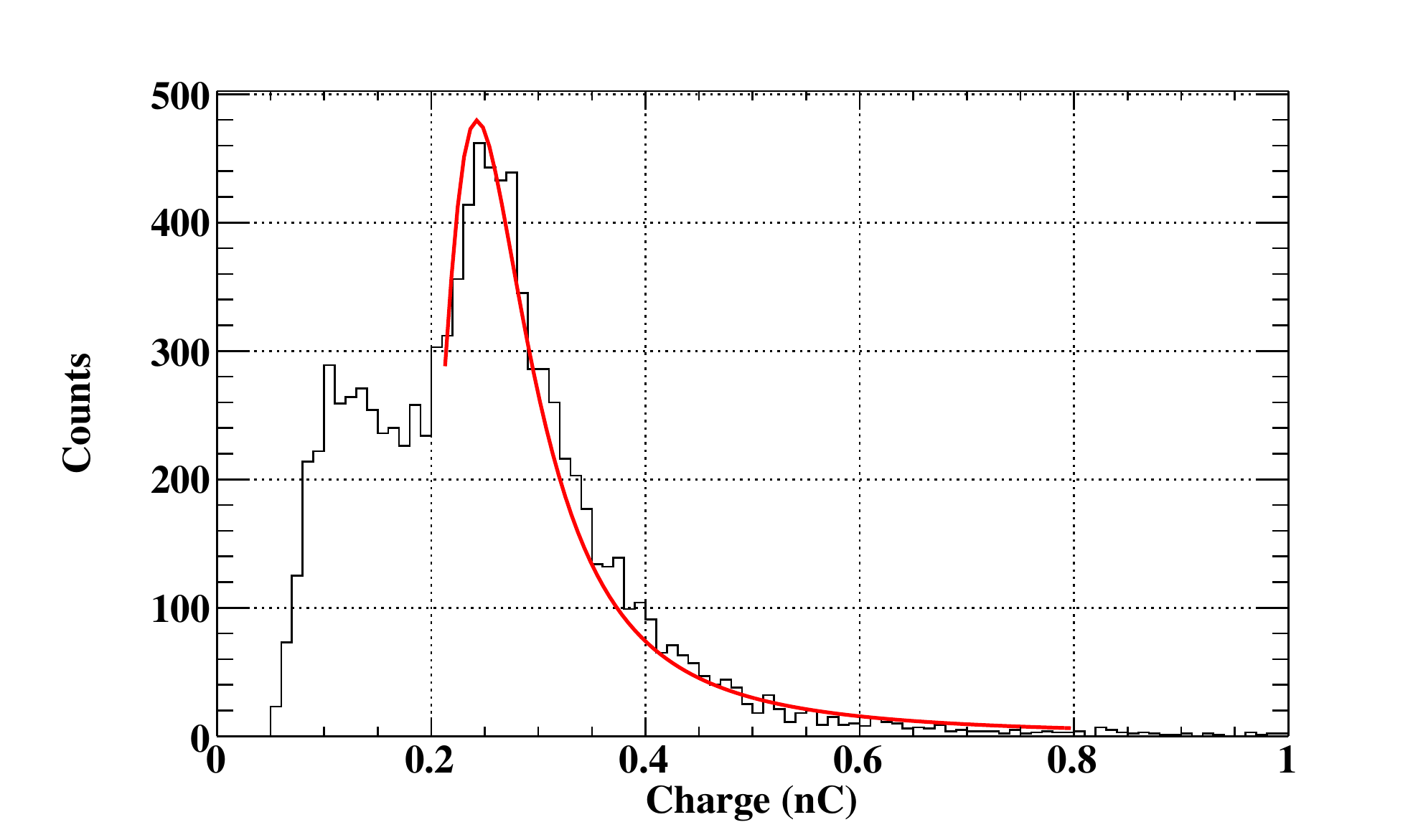}
\caption{Example of charge spectrum acquired with a TC bar, 
superimposed with a Landau function fit.}
\label{fig_landau}
\end{figure}

\begin{figure}[!ht]
\centering
\includegraphics[width=3.4in]{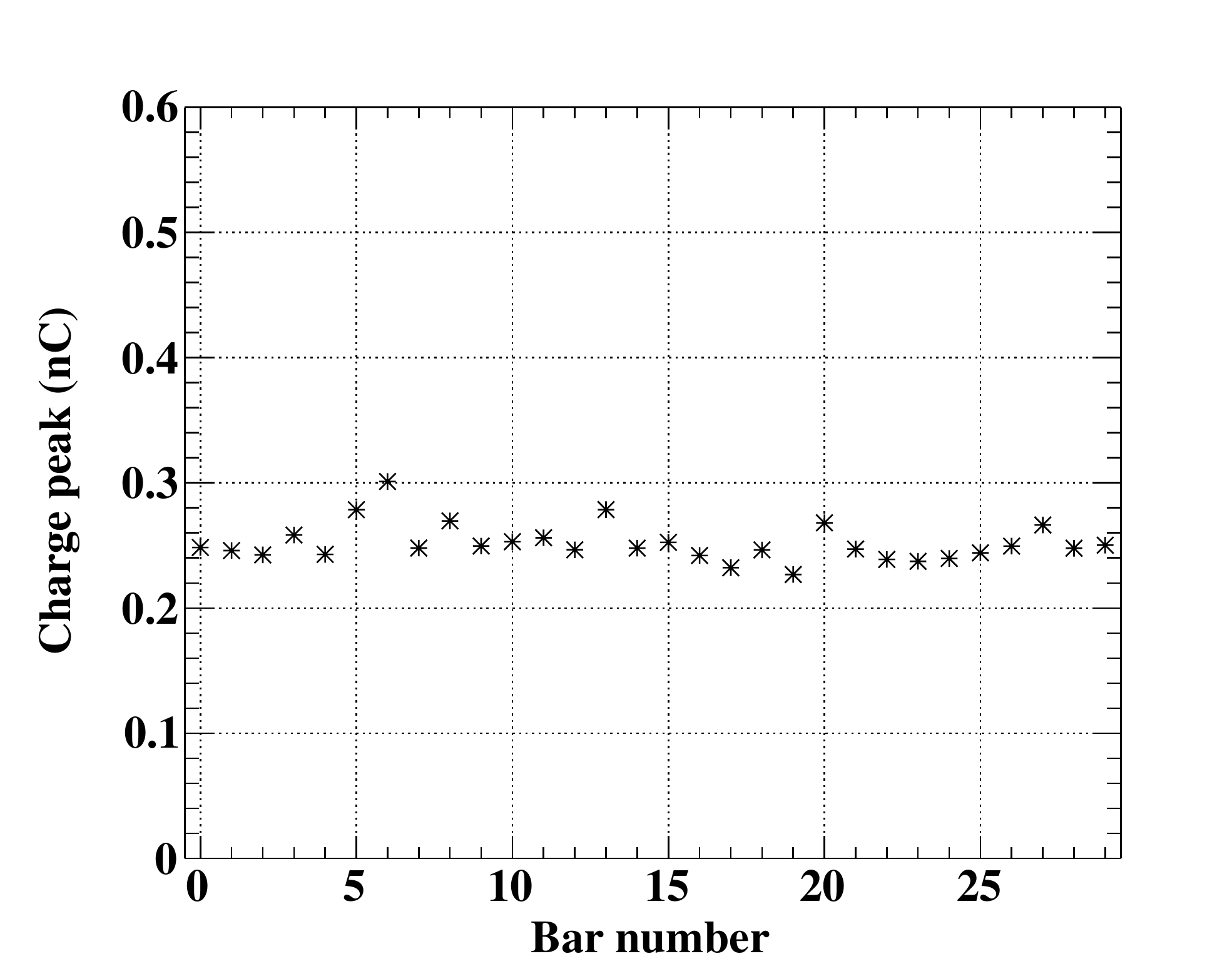}
\caption{Distribution of the Landau peak amplitudes versus the bar number.}
\label{landau_confronto}
\end{figure}

The PMT gain equalization procedure exploits cosmic rays hitting the TC:
due of the uniformity of cosmic ray hit distribution along the bars
the charge and amplitude spectra of the inner and outer PMTs are
marginally affected by geometrical effects.
It is therefore possible to tune the PMT gains acting on the bias high voltages
and equalize the peaks of the Landau distributions (Fig.\ref{fig_landau})
within $15\%$ maximum (Fig.\ref{landau_confronto}). A percent level
offline equalization is achieved
using suitable weights for each PMT in the trigger tables.
Due to the non-solenoidal configuration of the magnetic field,
inner and outer PMTs are subject to different field intensities and
orientations and therefore operate in very different working conditions.

\subsection{Time walk corrections}\label{sec_timewalk}
\begin{figure}[!ht]
\centering
\includegraphics[width=3.4in]{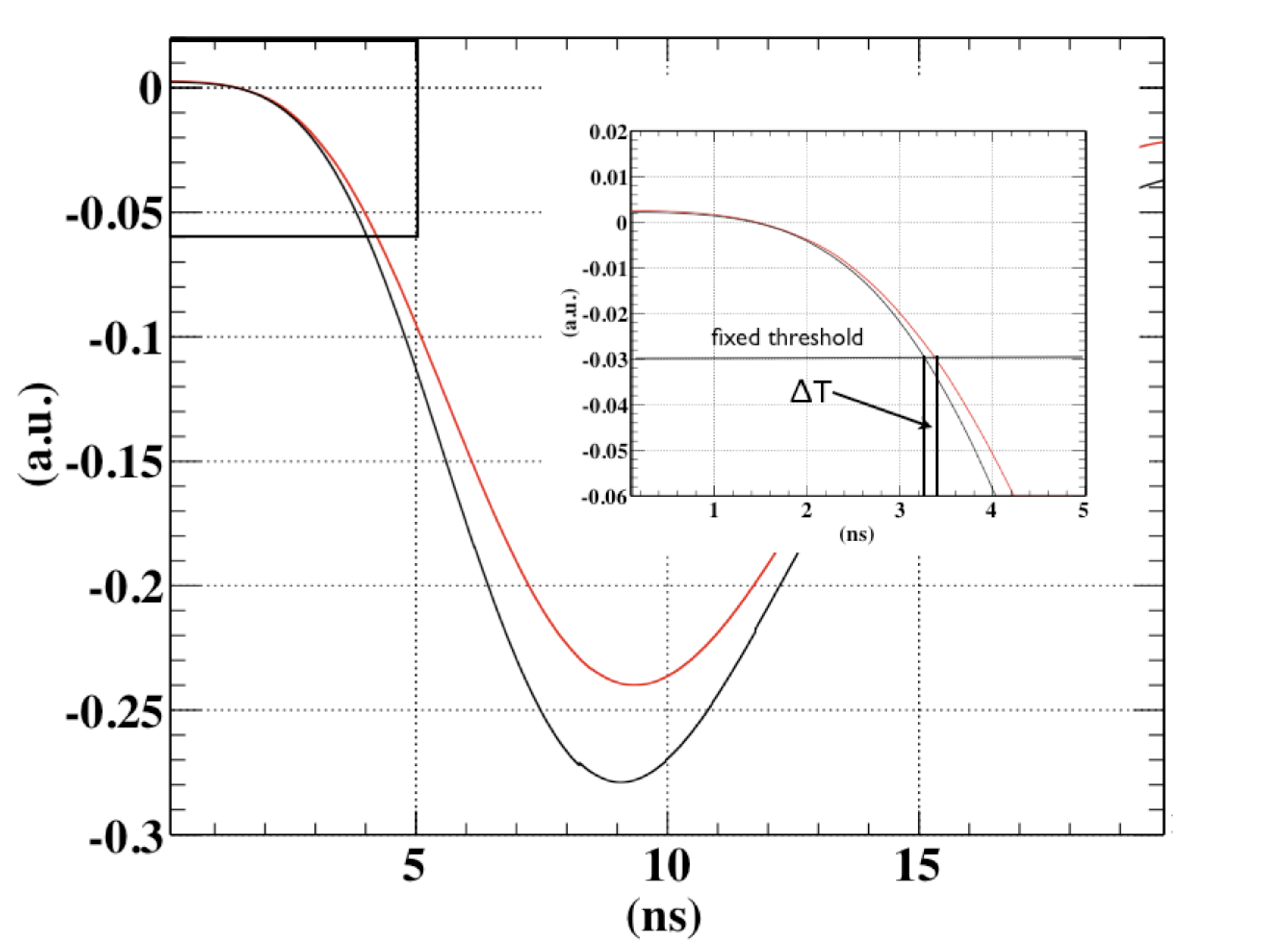}
\caption{Graphical representation of the Time Walk effect.}
\label{fig_timewalk}
\end{figure}

\begin{figure}[!ht]
\centering
\includegraphics[width=3in]{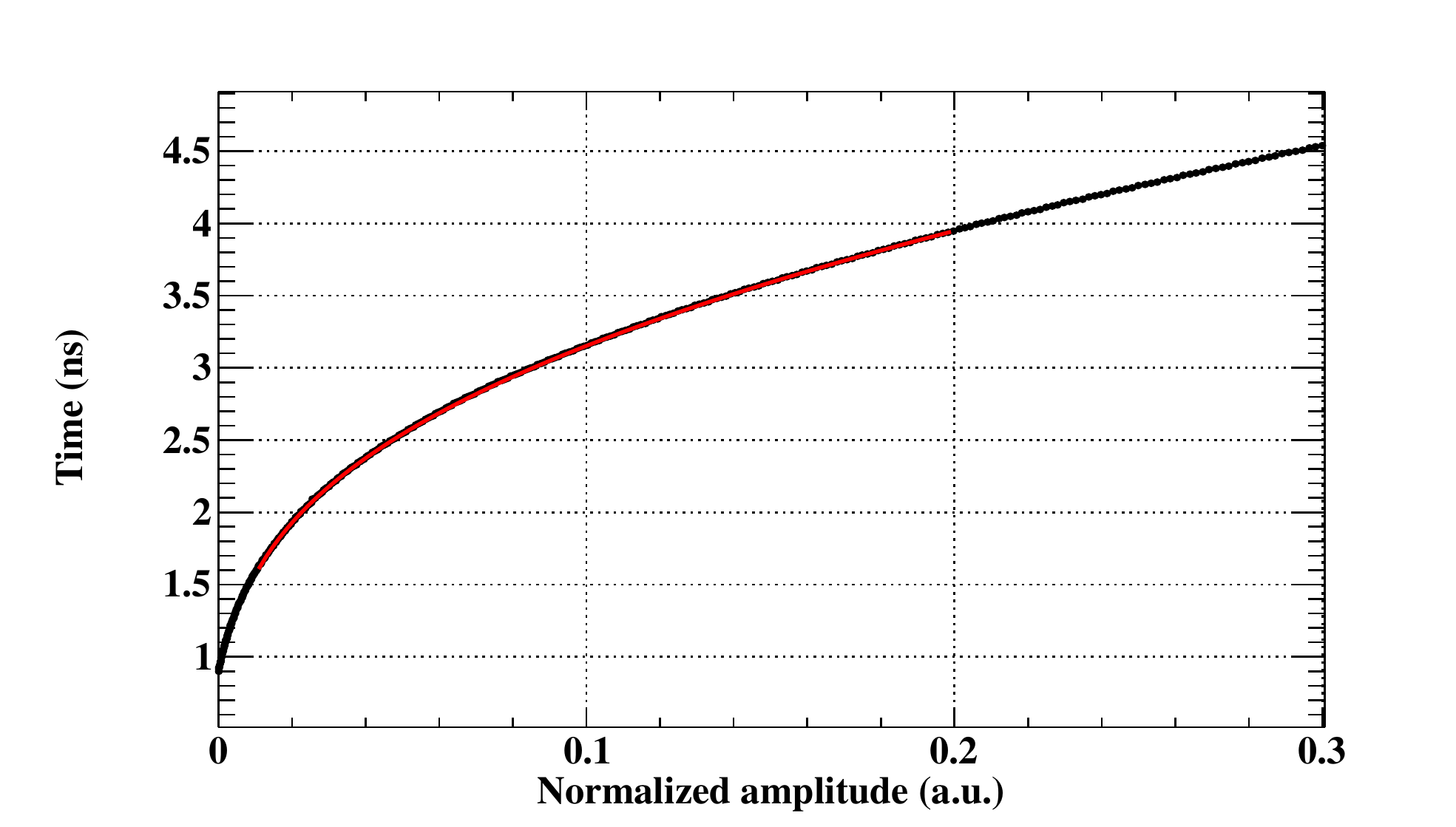}
\caption{Time delay versus amplitude relation with fit superimposed.}
\label{fig_time_amp}
\end{figure}

The Time Walk effect is the dependence of threshold crossing
delay time on pulse amplitude as showed in Fig.\ref{fig_timewalk}.
The analog waveforms sampled at $1.6\,\mathrm{GHz}$ are averaged and
interpolated separately for each PMT over many events to obtain a
template waveform, that can be used to evaluate the correction
required in time reconstruction: for each template pulse the leading
edge region is used to produce a plot of the delay time versus pulse
amplitude dependence that is fitted analytically with the function: 
\begin{equation}
TW=A+B\sqrt{x}+C\log{x}
\end{equation}
where x is the ratio between the low threshold value and the pulse
amplitude of the PMT and A, B, C are the fit parameters. As
shown in Fig.(\ref{fig_time_amp}), this function reproduces
outstandingly the experimental data. 

For each PMT, an event-by-event correction is applied
based on the pulse amplitude and the fitted coefficients.
This correction depends on the low level threshold (LLT)
value and the PMT pulse shape.  
The optimal LLT value balancing Time Walk immunity and noise rejection
was found evaluating the
timing resolution of each bar with dedicated tests performed using the
method described in \ref{sec_reso}.

\begin{figure}[!ht]
\centering
\includegraphics[width=3in]{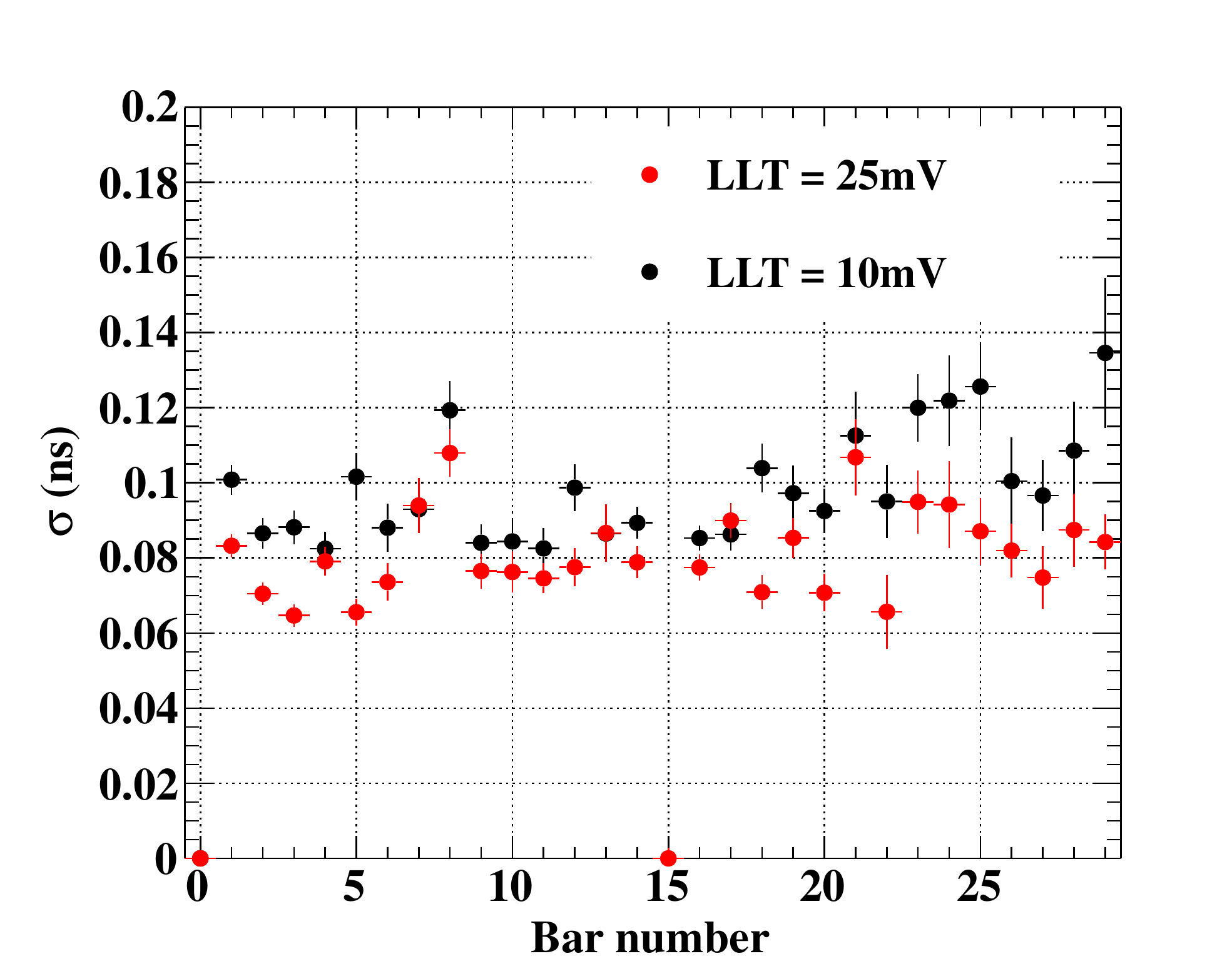}
\caption{Timing resolutions for different values of low level threshold for
double bar events.}
\label{fig_confronto_low}
\end{figure}

Fig.~\ref{fig_confronto_low} shows the comparison between the time
resolutions obtained with two different LLT values ($10\,\mathrm{mV}$
and $25\,\mathrm{mV}$) on the double bar sample.  
The difference between the time resolutions with the two threshold
showed is systematically favoring the higher value of
25~$\mathrm{mV}$. On the basis of
a complete scan of the time resolution (evaluated on the double bar
sample)  versus the low threshold in 5~$\mathrm{mV}$ steps from 5 to 
$35\,\mathrm{mV}$ the optimal value of LLT was found to be $25\,\mathrm{mV}$.

\subsection{Timing Counter offset correction}
Inter bar time offsets due to electronic chains are evaluated
using again cosmic rays.
All PMTs are equipped with Huber-S\"uhner Enviroflex\_400 signal
cables with equal length (10~$\mathrm{m}$) and with low loss
characteristics \cite{hubersuhner}. This minimizes the 
possibility of time offsets drifting during the data taking period
(several months) and, together with a
continuous monitoring of the relevant variables, ensures a stable
operation for our detector. For each bar, the time difference
distribution between inner and outer PMTs is acquired. Due to the
cosmic rays uniformity this distribution is expected to be flat and 
centered at zero; the mean values of these distributions give a
direct measure of the relative offsets between PMTs on each
scintillating bar. The time offsets between different bars and
TC and LXEC detectors are calculated using events from muon Michel
decays as well as different tools \cite{calibration_cw}.

\begin{figure}[!ht]
\centering
\includegraphics[width=3.7in]{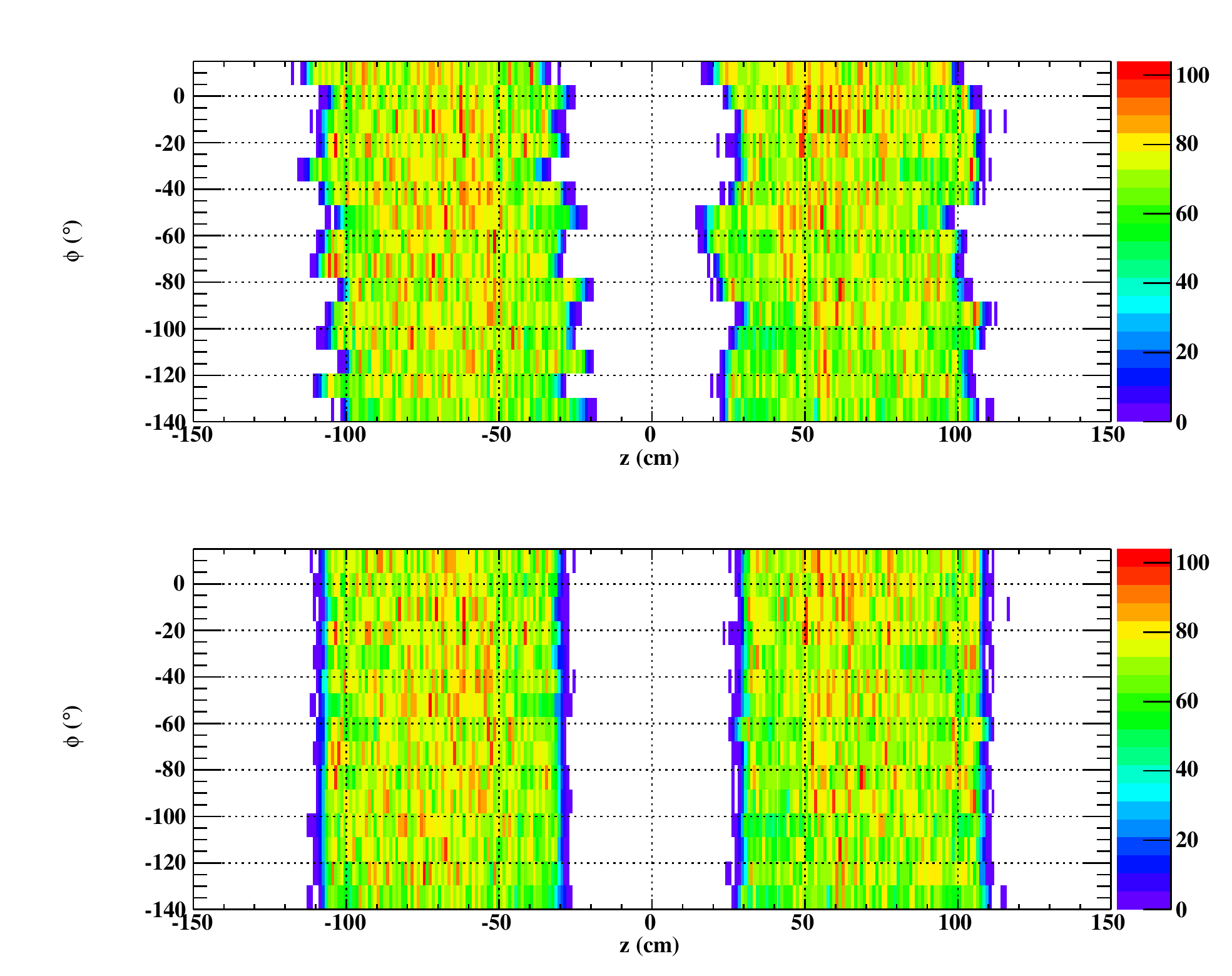}
\caption{Cosmic hit maps before and after bar time offset subtractions.}
\label{fig_hitmap}
\end{figure}

Fig.~\ref{fig_hitmap} shows the effect of accounting for bar offsets
on cosmic ray two-dimensional hit map. The shape of the detector hit
map becomes rectangular when the correct time offsets (lower panel)
are used and the $z$ reconstruction is 
reliable. This is to be compared to the upper panel, where the uncorrected
time differences are used for the impact point determination resulting in
a deformed hit map with respect to the actual detector shape.
Eq.~\ref{z_time} clearly implies that the width of
each distribution is directly proportional to $v_{\mathrm{eff}}$ 
in the corresponding bar.

The monitoring of the stability of the bar offset is achieved throughout the
full period of data taking with several tools both during the data
taking itself ({\em i.e.} using the Michel positrons to check the
hitmap) and with dedicated cosmic runs. Full calibration of time
offsets and TW coefficients is repeated before each year-long campaign
(to account for electronics and detector maintenance, rescaling of
gain factors).

\subsection{DTD calibration}\label{sec_DTD_eff}
The background rejection is a crucial point of the experiment,
together with the  maximum efficiency for the relevant signals: both
are related to the fast on-line event reconstruction performed at the
trigger level. Trigger position resolution is a key parameter for
direction match with the photon entering the LXEC and is evaluated to
be $\sigma(\Delta z)\sim 7.3\,\mathrm{cm}$ by comparing the
trigger-level reconstruction of impact point and the one obtained with
calibrated $\Delta t$. 

\begin{figure}[!ht]
\centering
\includegraphics[width=3.4in]{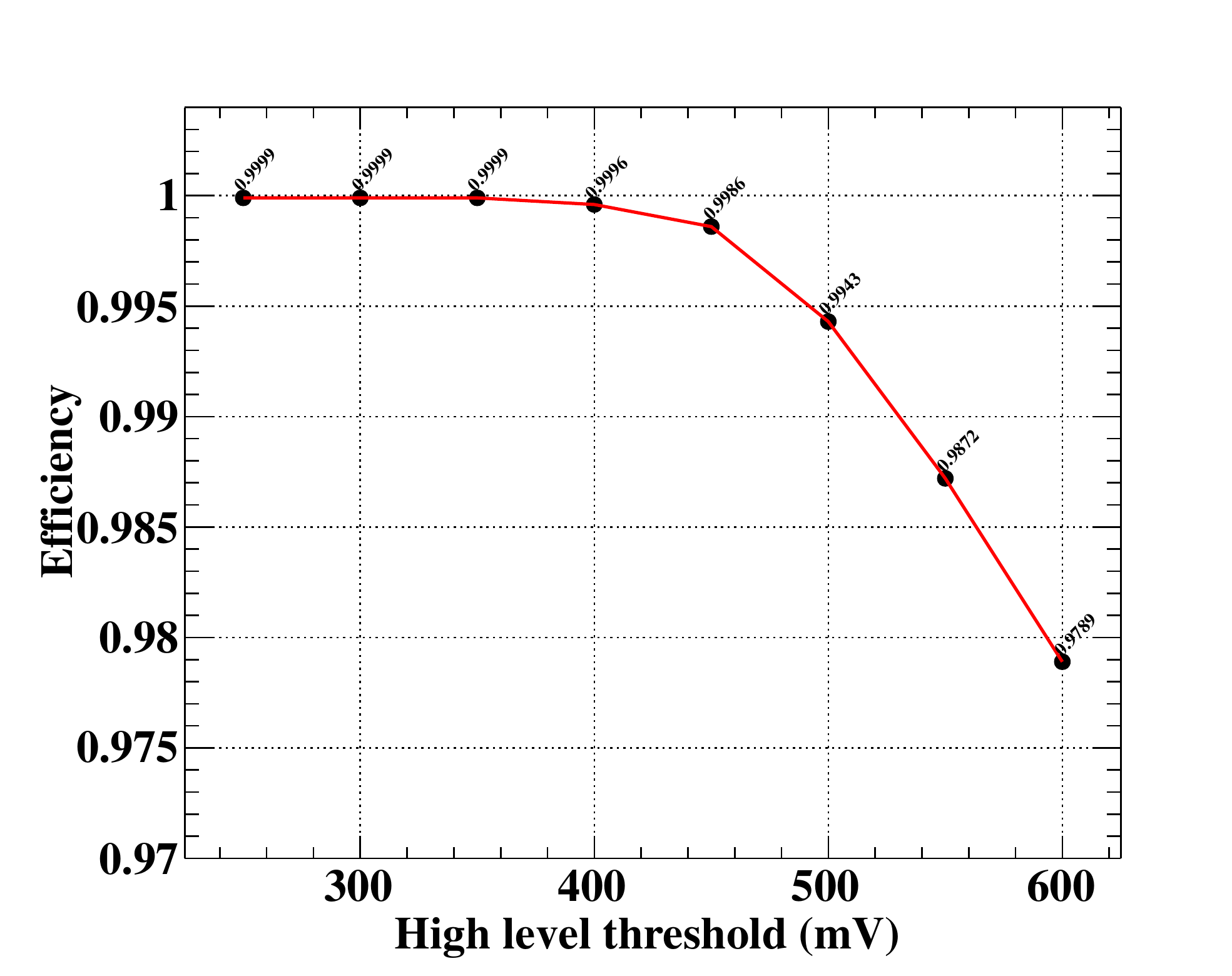}
\caption{DTD efficiency vs high threshold value. The efficiency
 plateau is reached for $HT \le 400\,{\mathrm mV}$.}
\label{fig_confronto_high}
\end{figure}

On the other hand, from the point of view of subsequent analysis, TC
trigger and DTD conditional efficiency study is motivated by
the request for all triggered events to fire a DTD output in order to
allow the best event reconstruction.
The relevant parameter is the threshold efficiency $\epsilon_{DTD}$
defined, for each bar, as the ratio between the number of events with
NIM signals and the total number of triggered events on the bar
itself, plotted versus the high level threshold. This one is set slightly
lower than the trigger one, in order to avoid acquiring events without
NIM pulse information. Fig.~\ref{fig_confronto_high} presents the
efficiency $\epsilon_{DTD}$ versus threshold level. Averaging over all
bars we obtain $\epsilon_{DTD} \ge 99.9\%$ for  High Threshold
$\mathrm{HLT} \le 400\,\mathrm{mV}$.  

\section{Resolutions}

\subsection{Timing resolution}\label{sec_reso}
\begin{figure}[!ht]
\centering
\includegraphics[width=3.4in]{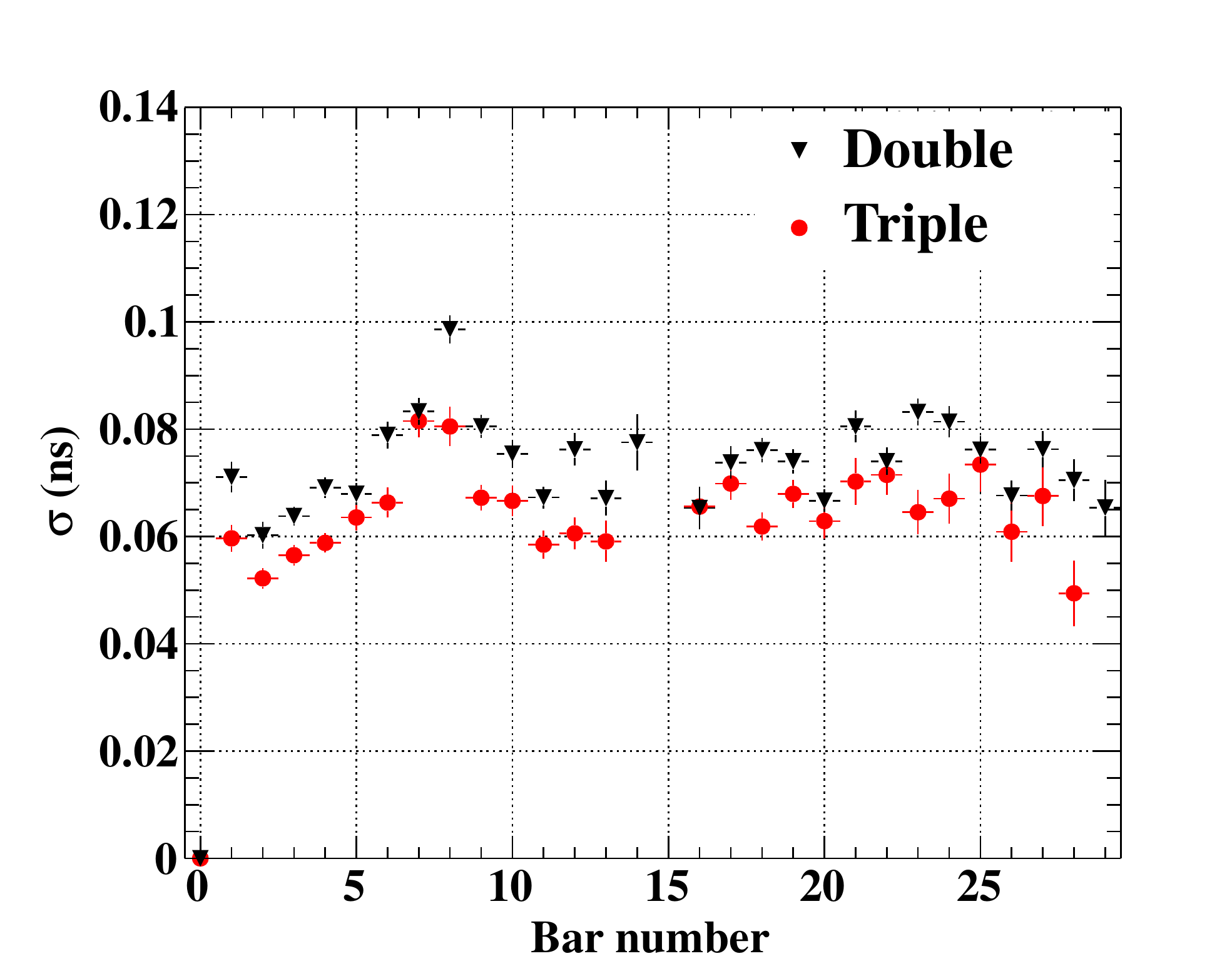}
\caption{Timing resolution on double (black markers) and triple (red
  markers) bar events, with LLT=$25~\mathrm{mV}$. Due to the different
  rates for two and three-bar events, the sample used here is
  different than the one in Fig.~\ref{fig_confronto_low}} 
\label{fig_reso}
\end{figure}

The timing resolution is extrapolated from multiple bars hit. On a
sample of two consecutive bars hit by a positron, the time difference
$\Delta T = T_{2} -T_{1}$ is studied. For each bar the impact time is
defined as in Eq.~\ref{tctime}. 
The time difference can be written as
\begin{equation} 
\sigma_{T} =\sqrt{(\sigma^2_{1} + \sigma^2_{2}) + \sigma^2_{track}}
\end{equation}
(where $\sigma_{i}$ are the single bar time resolutions and
$\sigma_{track}$ the contribution from the track length spread. Under
the hypothesis that the two bars have similar resolutions $\sigma_{1}
\approx \sigma_{2} $ and neglecting the term due to 
track propagation, the single bar resolution is estimated from 
the spread of the $\Delta T$ distributions divided by $\sqrt{2}$. More
precisely, we perform a fit using a Gaussian plus second order
polynomial function to take into account tails due to  spurious hits
(less than a few \% of the total). In order to keep a conservative
estimate useful also for long-term monitoring, we apply 
looser cuts with respect to the final analysis: for example, by requesting 
a track reconstructed in the DCH, the polynomial
tail disappears and the core resolution remains unchanged.
This method slightly overestimate the timing resolution since the
track length spread term is not corrected for. Using triple bar
events and evaluating the quantity:
\begin{equation}
\Delta T = T_{2} - \frac{T_1+T_3}{2}
\end{equation}
the effect of the different path length between bars can be rejected
at the first order, resulting in a better estimate of timing
resolution, as showed in Fig.~\ref{fig_reso} where the red and black
markers represent respectively
the resolutions obtained in triple and double bar samples.
However the triple sample has significantly smaller statistics so the
double bar sample is kept as the reference tool for checking the detector
performances. The resolutions reported in Fig.~\ref{fig_reso} are
obtained from a different sample than those shown in
Fig.~\ref{fig_confronto_low}: this was because of the insufficient
number of triple bar events in the latter one. However the two data
sets show very similar behavior pointing out a stable operation of the
detector. 

A certain degradation of the timing performance between the Beam Test
($\sigma_t = 40-50~\mathrm{ps}$) and the Physics Run configuration
($\sigma_t = 60-80~\mathrm{ps}$) has been observed. A few factors
have contributed to this effect: the need for a lower PMT gain in order to
withstand the high rate and match the dynamic range of the
DAQ/electronics chain; a slightly higher value for the low threshold
due to additional noise from the surrounding environment; and intrinsic
uncertainties in both double and triple bar estimates. A deeper study
on alternative schemes for extracting the time information is under
way in order to improve the detector performance.

\subsection{Impact point resolution}
\begin{figure}[!ht]
\centering
\includegraphics[width=3.4in]{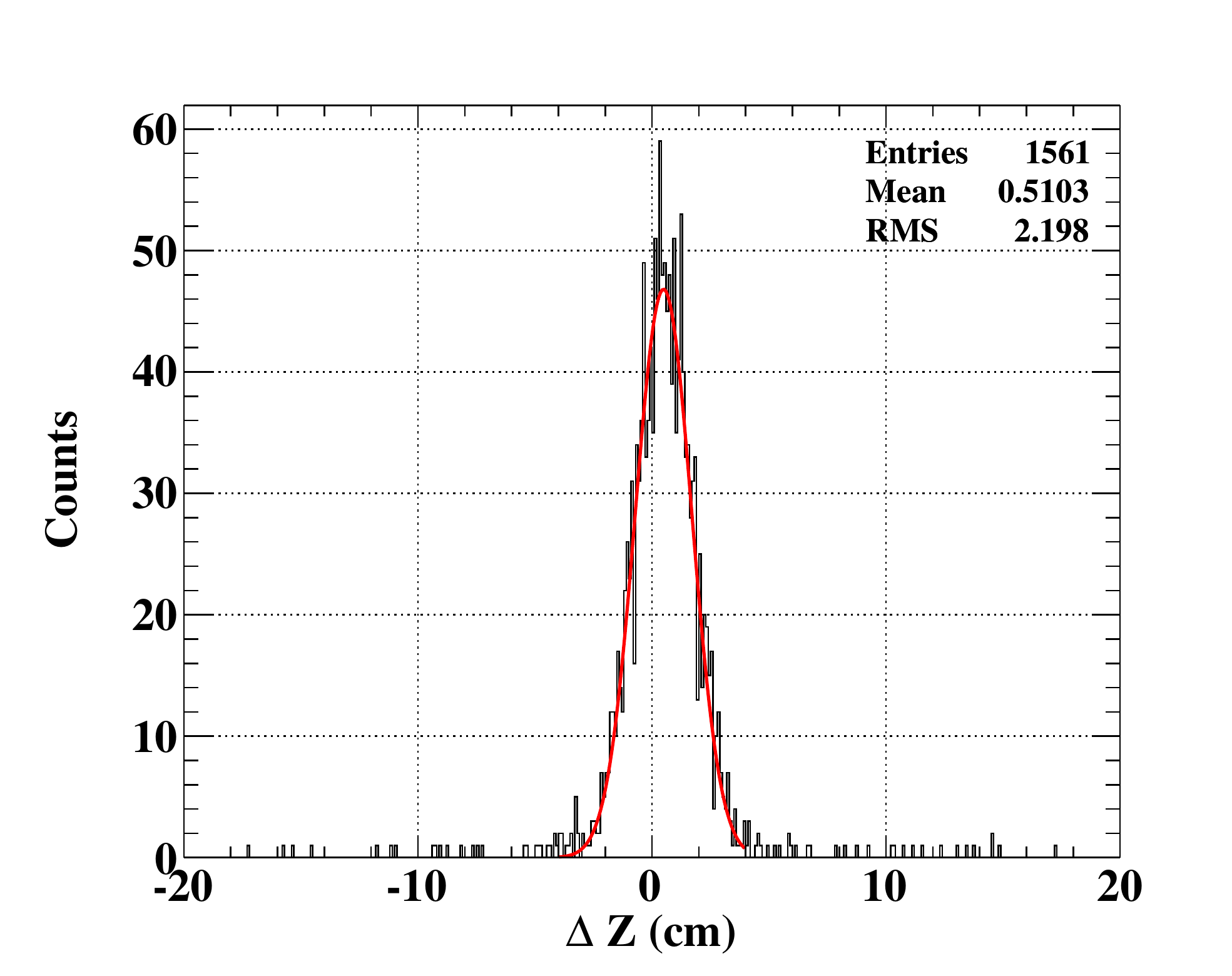}
\caption{$z$ resolution on triple bar events with a Gaussian fit superimposed.}
\label{fig_z_reso}
\end{figure}

From the triple bar events it is also possible to estimate the resolution 
on the z coordinate of the impact point, substituting the
reconstructed time with the reconstructed z coordinate:
\begin{equation}
\Delta z = z_{2}-\frac{z_{1}+z_{3}}{2}
\end{equation}
where $i=1,2,3$ is the bar index.
The position resolution obtained from the triple bar sample is 
$1.1\,\mathrm{cm} < \sigma_{z} <
2.0\,\mathrm{cm}$. Fig.\ref{fig_z_reso} shows an example of a 
$\Delta z$  distribution.

\section{Conclusion}
The TC is a cutting edge detector in the field of high resolution
timing measurements. 
It demonstrated a timing resolution as good as $40\,\mathrm{ps}$ in
dedicated beam tests and $60\,\mathrm{ps} < \sigma(t) <
80\,\mathrm{ps}$ under real conditions in the MEG experiment together
with a precise measurement of the impact point with
$\sigma(z)<2\,\mathrm{cm}$. The TC fast response satisfies all the
requirements of a trigger designed to select events on-line in a huge
background. The TC has been commissioned and is currently operating as
part of the MEG detector.

\section*{Acknowledgments}

We want to acknowledge Dr.~M.~Ribeiro Gomes for her valuable help in
revising the 
manuscript and  
the mechanical and electronics workshops at  INFN Sections of Genova and Pavia.

\ifCLASSOPTIONcaptionsoff
  \newpage
\fi


\bibliographystyle{IEEEtran}
%

\bibliography{megtc}

%








\end{document}